\documentclass[aps,prresearch,reprint,twocolumn,superscriptaddress,floatfix,nofootinbib,longbibliography]{revtex4-1}
\usepackage{graphicx,amsmath,amsfonts,amssymb,amsthm,xr}
\usepackage{epsfig,amsmath,amssymb,color,dsfont,upgreek,physics}
\usepackage{mathrsfs}
\usepackage{mathtools}
\usepackage{bbold}
\usepackage{comment}
\usepackage{float}
\usepackage{soul}

\usepackage[bookmarks=true,colorlinks,linkcolor=OrangeRed,urlcolor=NavyBlue,citecolor=RoyalBlue]{hyperref}
\usepackage[dvipsnames]{xcolor}
\usepackage{orcidlink}

\definecolor{mygold}{rgb}{0.93,0.69,0.13}

\definecolor{mypurple}{rgb}{0.49,0.18,0.56}

\definecolor{mygreen}{rgb}{0,0.5,0}

\definecolor{mygreen}{rgb}{0,0.5,0}
\definecolor{myred}{rgb}{0.7,0,0}
\definecolor{myblue}{rgb}{0,0,0.5}

\begin{document}
\title{Anatomy of Dynamical Quantum Phase Transitions}
\author{Maarten Van Damme${}^{\orcidlink{0000-0002-1558-0568}}$}
\thanks{M.V.D.~and J.-Y.D.~contributed equally to this work}
\affiliation{Department of Physics and Astronomy, University of Ghent, Krijgslaan 281, 9000 Gent, Belgium}
\author{Jean-Yves Desaules${}^{\orcidlink{0000-0002-3749-6375}}$}
\thanks{M.V.D.~and J.-Y.D.~contributed equally to this work}
\affiliation{School of Physics and Astronomy, University of Leeds, Leeds LS2 9JT, UK}
\author{Zlatko Papi\'c${}^{\orcidlink{0000-0002-8451-2235}}$}
\affiliation{School of Physics and Astronomy, University of Leeds, Leeds LS2 9JT, UK}
\author{Jad C.~Halimeh${}^{\orcidlink{0000-0002-0659-7990}}$}
\email{jad.halimeh@physik.lmu.de}
\affiliation{Department of Physics and Arnold Sommerfeld Center for Theoretical Physics (ASC), Ludwig-Maximilians-Universit\"at M\"unchen, Theresienstra\ss e 37, D-80333 M\"unchen, Germany}
\affiliation{Munich Center for Quantum Science and Technology (MCQST), Schellingstra\ss e 4, D-80799 M\"unchen, Germany}

\begin{abstract}
Global quenches of quantum many-body models can give rise to periodic dynamical quantum phase transitions (DQPTs) directly connected to the zeros of a Landau order parameter (OP). The associated dynamics has been argued to bear close resemblance to Rabi oscillations characteristic of two-level systems. Here, we address the question of whether this DQPT behavior is merely a manifestation of the limit of an effective two-level system or if it can arise as part of a more complex dynamics. We focus on quantum many-body scarring as a useful toy model allowing us to naturally study state transfer in an otherwise chaotic system. We find that a DQPT signals a change in the dominant contribution to the wave function in the degenerate initial-state manifold, with a direct relation to an OP zero only in the special case of occurring at the midpoint of an evenly degenerate manifold. Our work generalizes previous results and reveals that, in general, periodic DQPTs comprise complex many-body dynamics fundamentally beyond that of two-level systems.
\end{abstract}

\date{\today}
\maketitle

\section{Introduction}

One of the central goals of far-from-equilibrium quantum many-body physics is the understanding of dynamical quantum universality classes, the pursuit of which has led to the introduction of several concepts of dynamical phase transitions \cite{Zvyagin2016,Mori_review,Heyl_review,Marino_review}. Extending the concept of spontaneous symmetry breaking in equilibrium, one concept of dynamical phase transitions is characterized by the order parameter (OP) of the long-time steady state following a quench in a control parameter after starting in an ordered (symmetry-broken) initial state \cite{Sciolla2010,Sciolla2011,Halimeh2017a}. The critical value of the quench parameter separates a symmetry-broken from a symmetric steady state. In addition to this \textit{Landau type} of dynamical phase transitions, another concept has been introduced, known as \textit{dynamical quantum phase transitions} (DQPTs), that relies on a connection to thermal phase transitions \cite{Heyl2013}. It is based on recognizing that the overlap $\braket{\psi_0}{\psi(t)}$ between the initial state $\ket{\psi_0}$ and the time-evolved wave function $\ket{\psi(t)}=e^{-i\hat{H}t}\ket{\psi_0}$, with $\hat{H}$ the quench Hamiltonian, is a boundary partition function with complexified time $it$ representing inverse temperature. Equivalently, the \textit{return rate}, ${-}\lim_{L\to\infty}L^{-1}\ln\lvert\braket{\psi_0}{\psi(t)}\rvert^2$, with $L$ the system size, becomes a dynamical analog of the thermal free energy, with a DQPT formally defined as a nonanalyticity in it at a \textit{critical time} $t_\text{c}$.

\begin{figure}[t!]
	\centering
 	\includegraphics[width=\linewidth]{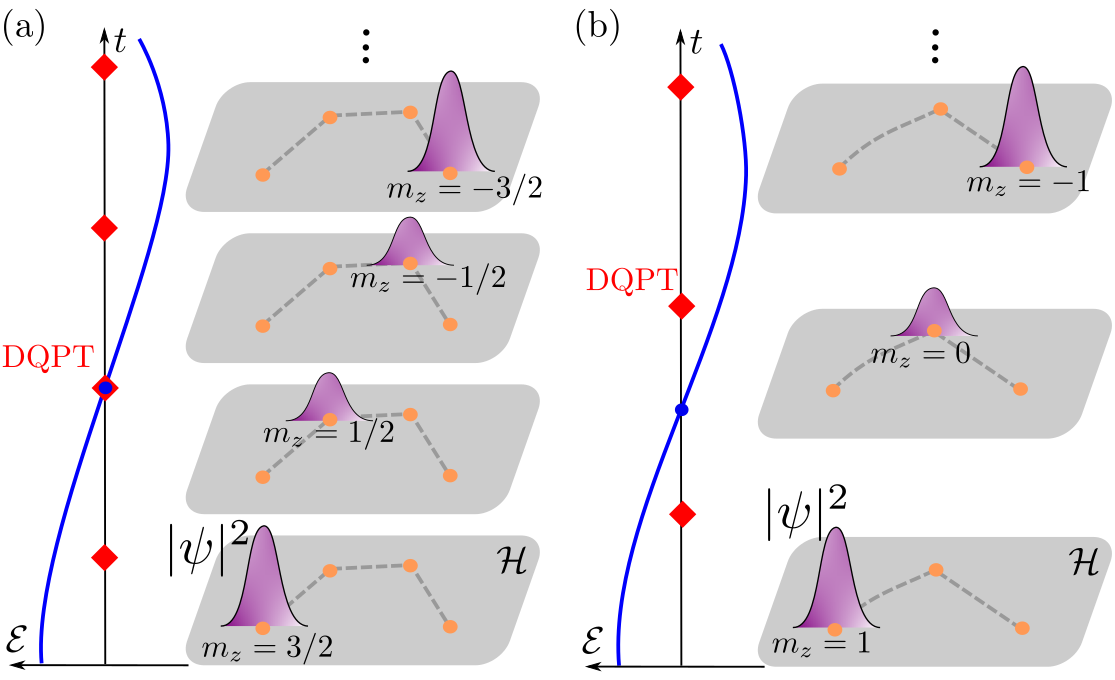}
	\caption{(Color online). Schematic of the main conclusions of our work. During ``state-transfer'' quench dynamics in the spin-$S$ $\mathrm{U}(1)$ QLM initialized in an \textit{extreme vacuum}, DQPTs (red diamonds) arise in the return rate (RR) when there is a shift in wave function-overlap dominance between components of the degenerate vacuum manifold, where each component vacuum is denoted by $m_z{\in}\{-S,\ldots,S\}$ (orange dots) in the total Hilbert space $\mathcal{H}$. (a) For $S{=}3/2$ we find a DQPT occurring at the same time as a zero in the OP $\mathcal{E}(t)$ (blue dot). This DQPT signals state transfer between the intermediate vacua $m_z{=}{\pm}1/2$. Other DQPTs show no such connection. This can be generalized for any half-integer $S$. (b) For $S{=}1$, a DQPT and an OP zero do not occur simultaneously. Instead, the OP zero is halfway between two consecutive DQPTs that signal state transfer to and away from the middle vacuum $m_z{=}0$. This holds for all integer $S$. This picture generalizes previous results and highlights the complex many-body dynamics comprising DQPTs.}
	\label{fig:schematic}
\end{figure}

In a wide variety of quantum many-body models hosting a global symmetry,  DQPTs in the wake of sufficiently large quenches starting in the ordered phase have been shown to be directly connected to zeros of the OP dynamics \cite{Heyl2014,Halimeh2017,Homrighausen2017,Zunkovic2018,Huang2019,Zache2019,Halimeh2020quasiparticle,Hashizume2022}. In the seminal work introducing DQPTs, the model showing this behavior is the integrable transverse-field Ising chain (TFIC) \cite{Heyl2013}.  This model can be solved exactly by mapping it to a two-band free fermionic model using a Jordan--Wigner transformation, where each disconnected momentum sector is a two-level system. For nonintegrable models, e.g., the long-range interacting TFIC or two-dimensional quantum Ising models, the direct connection between DQPTs and OP zeros occurs for large quenches, where the transverse-field strength is much larger than the coupling constant, and which can be considered perturbatively close to classical precession in two-level systems during the short timescales where this behavior is prominent \cite{Halimeh2017,Zunkovic2018,Halimeh2020quasiparticle}. Indeed, for intermediate quenches where this perturbative treatment is no longer valid, this connection breaks down \cite{Sirker2014,Vajna2014,Karrasch2013}, and for small quenches within the ordered phase, \textit{anomalous} DQPTs can appear even without any OP zeros occurring over all investigated timescales \cite{Halimeh2017,Homrighausen2017,Corps2022}. As such, it has been argued that the periodic DQPT behavior seen for large quenches may be a mere manifestation of effective two-level system dynamics \cite{Zakrzewski2022}. 

Here, we address this argument by investigating DQPTs in models exhibiting ``state transfer''---a dynamical process where the wave function evolves cyclically between the states of a given manifold---due to quantum many-body scarring~\cite{Serbyn2021, MoudgalyaReview, ChandranReview}. Such models possess a small number of anomalous eigenstates throughout their spectrum, leading to oscillatory dynamics from a few specific states in an otherwise thermalizing system.  We will focus on a formulation  of the lattice Schwinger model~\cite{Schwinger1962} known as the spin-$S$ $\mathrm{U}(1)$ quantum link model (QLM) \cite{Chandrasekharan1997,Wiese_review}. This model has been shown to exhibit quantum many-body scarring for massless quenches starting in the maximal-flux (extreme) vacua for a wide range of spin values~\cite{Surace2020,Desaules2022a}. We show in these models that periodic DQPTs arise within complex many-body dynamics that is beyond two-level systems, and where a DQPT signals state transfer from one vacuum to another within a $(2S{+}1)$-fold degenerate vacuum manifold; see Fig.~\ref{fig:schematic}. We find no direct connection between DQPTs and OP zeros for integer $S$. For half-integer $S$, a DQPT is directly connected to an OP zero only when the DQPT signals a transfer between intermediate minimal-flux vacua of opposite flux sign. We further show that models where DQPT behavior resembles two-level system dynamics are a special case of our general picture.

\section{Model}
We consider the spin-$S$ $\mathrm{U}(1)$ QLM, given by the Hamiltonian \cite{Chandrasekharan1997,Wiese_review,Kasper2017}
\begin{align}\nonumber
    \hat{H}=\sum_{j=1}^L\bigg[&\frac{J}{2\sqrt{S(S+1)}}\big(\hat{\sigma}^-_j\hat{s}^+_{j,j+1}\hat{\sigma}^-_{j+1}+\text{H.c.}\big)\\\label{eq:QLM}
    &+\mu\hat{\sigma}^z_j+\frac{\kappa^2}{2}\big(\hat{s}^z_{j,j+1}\big)^2\bigg],
\end{align}
where we have adopted particle-hole and Jordan--Wigner transformations \cite{Hauke2013,Yang2016}. The Pauli operator $\hat{\sigma}^z_j$ describes the matter occupation on site $j$ with mass $\mu$, and the spin-$S$ operators $\hat{s}^+_{j,j+1}{/}\sqrt{S(S+1)}$ and $\hat{s}^z_{j,j+1}$ represent the gauge and electric fields, respectively, residing on the link between sites $j$ and $j+1$. The tunneling constant is $J$, which we shall set to unity as the overall energy scale, $\kappa$ is the gauge-coupling strength, and $L$ is the number of sites.

The generator of the $\mathrm{U}(1)$ gauge symmetry of Hamiltonian~\eqref{eq:QLM} is
\begin{align}\label{eq:Gj}
    \hat{G}_j=(-1)^j\bigg(\hat{s}^z_{j-1,j}+\hat{s}^z_{j,j+1}+\frac{\hat{\sigma}^z_j+\hat{\mathds{1}}}{2}\bigg),
\end{align}
and gauge-invariant states $\ket{\phi}$ satisfy $\hat{G}_j\ket{\phi}{=}g_j\ket{\phi},\,\forall j$, where $g_j/(-1)^j{\in}\{{-}2S,\ldots,2S{+}1\}$. We will work in the \textit{physical} sector $g_j{=}0,\,\forall j$.

A building block of the $\mathrm{U}(1)$ QLM has been experimentally realized for $S{\to}\infty$ in a cold-atom setup \cite{Mil2020}. Large-scale implementations of the spin-$1/2$ $\mathrm{U}(1)$ QLM on a Bose--Hubbard superlattice have been employed to observe gauge invariance \cite{Yang2020} and thermalization dynamics \cite{Zhou2021}.

\section{Quench dynamics}
We now present time-evolution results obtained through the infinite matrix product state (iMPS) technique based on the time-dependent variational principle \cite{Haegeman2011,Haegeman2013,Haegeman2016,Vanderstraeten2019}. This technique works directly in the thermodynamic limit, and also allows us to directly detect DQPTs as level crossings between the logarithms of the eigenvalues of the MPS transfer matrix \cite{Zauner2015,Halimeh2017,Zauner2017}, without any need for finite-size scaling with $L$. DQPTs for gauge theories have already been studied in the context of the spin-$1/2$ $\mathrm{U}(1)$ QLM \cite{Huang2019,Pedersen2021,Jensen2022}, and also for $S{\geq}1/2$ \cite{VanDamme2022} as well as  in the Schwinger model \cite{Zache2019,Mueller2022DQPT}, but not in the context of quantum many-body scarring. For the most stringent calculations that we have performed in iMPS for this work, we find convergence for a maximal bond dimension of $550$ and a time-step of $0.0005/J$.

We are interested in the dynamics of the experimentally relevant return rate (RR)
\begin{subequations}\label{eq:RR_top}
\begin{align}\label{eq:RR}
    \lambda(t)&=\min_{m_z}\big\{\lambda_{m_z}(t)\big\},\\\label{eq:RRcomponent}
    \lambda_{m_z}(t)&=-\lim_{L\to\infty}\frac{1}{L}\ln\big\lvert\braket{\psi_0^{m_z}}{\psi(t)}\big\rvert^2,
\end{align}
\end{subequations}
which has recently been used to identify DQPTs in a trapped-ion experiment \cite{Jurcevic2017}. Here, $\big\{\ket{\psi_0^{m_z}}\big\}$ is the set of vacua with $m_z{\in}\big\{{-}S,\ldots,S\big\}$, which are the $(2S{+}1)$-fold degenerate ground states of Hamiltonian~\eqref{eq:QLM} for $\kappa/J{=}0$ and $\mu/J{\to}\infty$. We call a vacuum \textit{extreme} when $\lvert m_z\rvert{=}S$, and \textit{intermediate} when $\lvert m_z\rvert{<}S$. We can represent a vacuum state on the two-site two-link unit cell as $\ket{\psi_0^{m_z}}{=}\ket{-1,m_z,-1,-m_z}$, indicating the eigenvalue $-1$ of $\hat{\sigma}^z_j$ at each (empty) matter site and the eigenvalue $m_z$ of $\hat{s}^z_{j,j+1}$ on each link, but we will often refer to this vacuum simply by its value of $m_z$. A detailed discussion of how to calculate the RR in iMPS can be found in Ref.~\cite{Zauner2017}.

The system is initialized in the \textit{extreme} vacuum $\ket{\psi_0^S}$ and subsequently quenched with the QLM Hamiltonian~\eqref{eq:QLM} with $\mu{/}J{=}\kappa{/}J{=}0$, yielding the time-evolved wave function $\ket{\psi(t)}{=}e^{-i\hat{H}t}\ket{\psi_0^S}$. Furthermore, we will also calculate in iMPS the quench dynamics of the electric flux and chiral condensate,
\begin{subequations}\label{eq:LocObs}
\begin{align}\label{eq:flux}
    \mathcal{E}(t)&=\lim_{L\to\infty}\frac{1}{L}\sum_{j=1}^L(-1)^{j+1}\bra{\psi(t)}\hat{s}^z_{j,j+1}\ket{\psi(t)},\\\label{eq:CC}
    n(t)&=\frac{1}{2}+\lim_{L\to\infty}\frac{1}{2L}\sum_{j=1}^L\bra{\psi(t)}\hat{\sigma}^z_j\ket{\psi(t)}.
\end{align}
\end{subequations}
The electric flux is an OP associated with the global $\mathbb{Z}_2$ symmetry of Hamiltonian~\eqref{eq:QLM} \cite{Coleman1976}.

\begin{figure}[t!]
	\centering
	\includegraphics[width=0.48\textwidth]{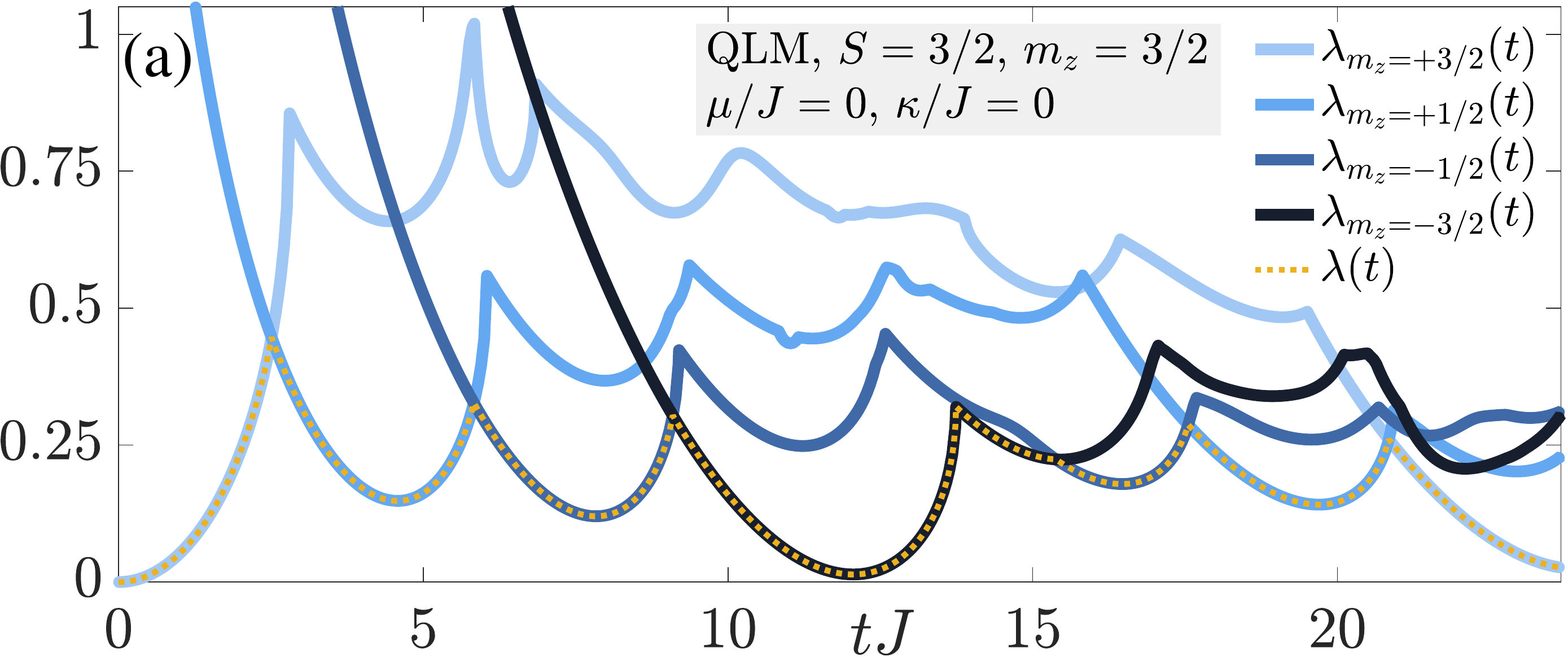}\\
	\vspace{1.1mm}
	\includegraphics[width=0.48\textwidth]{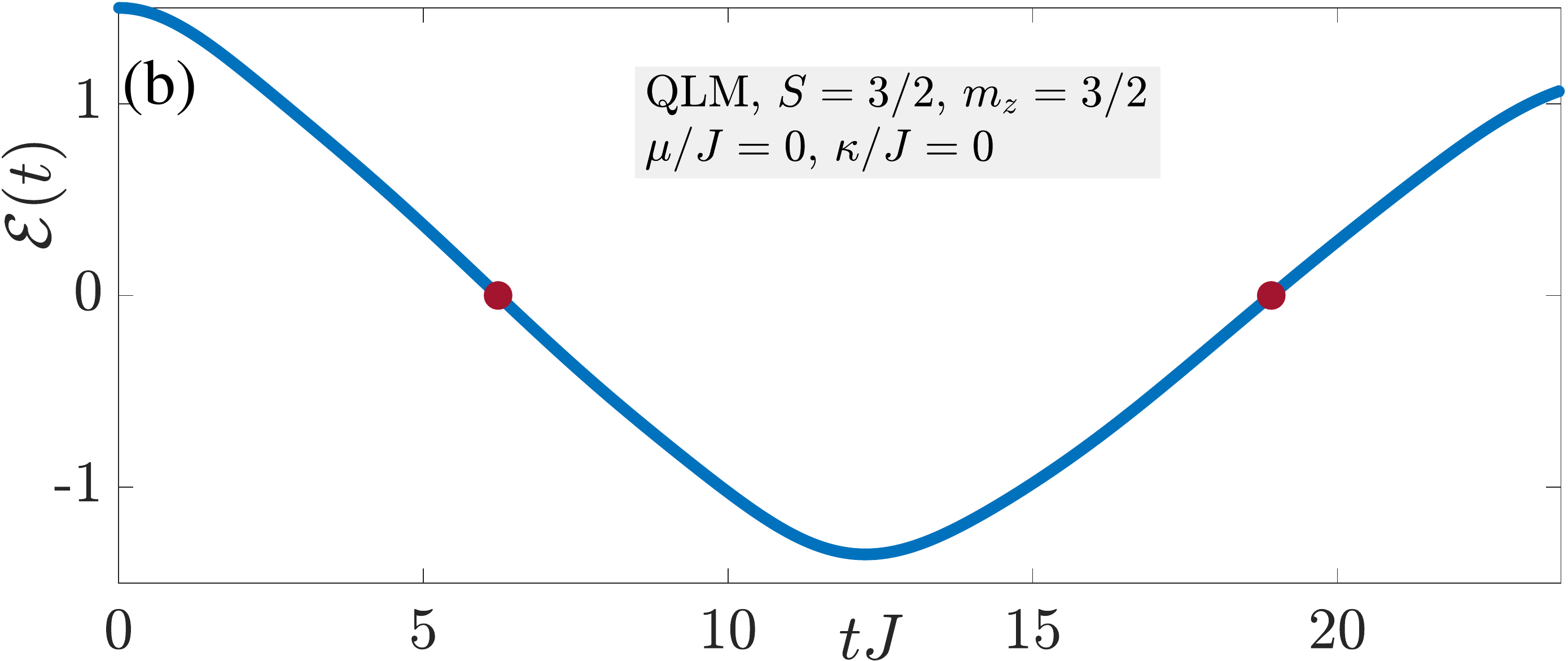}\\
	\vspace{1.1mm}
	\includegraphics[width=0.48\textwidth]{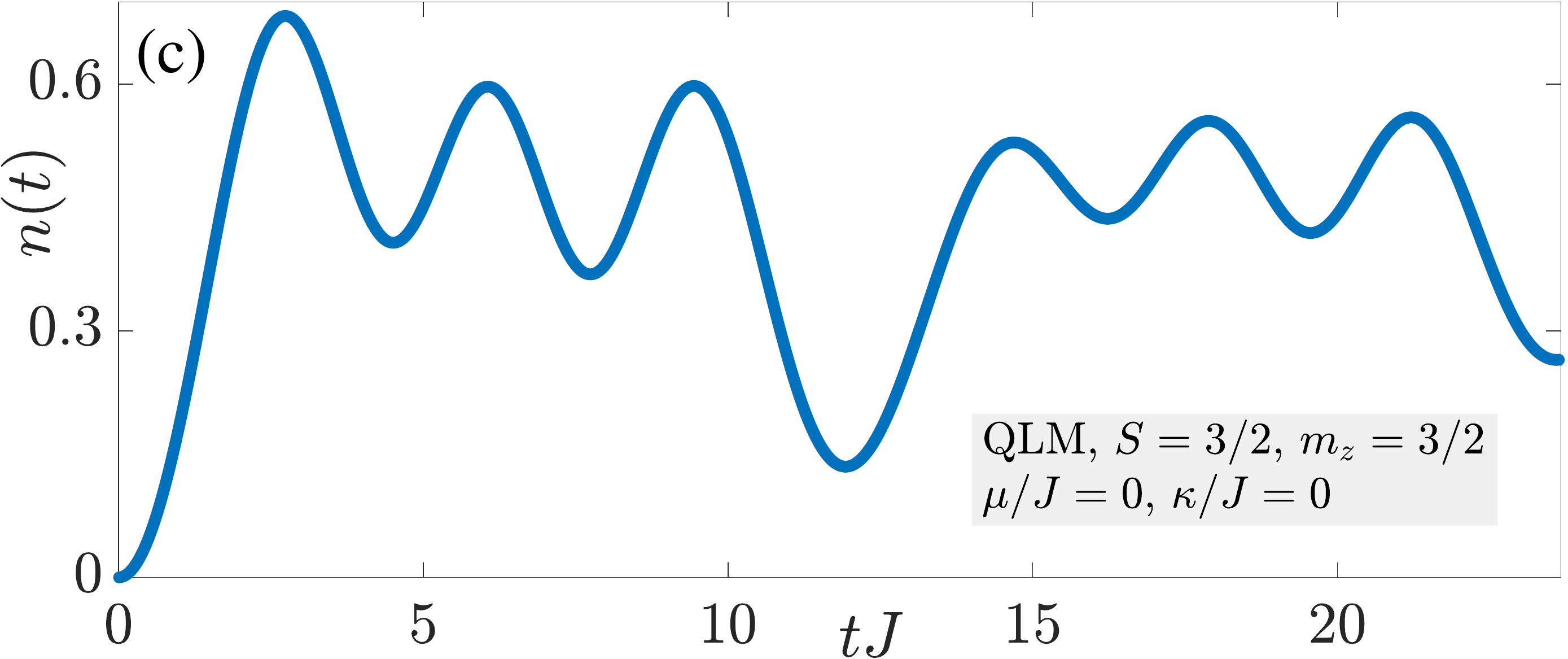}
	\caption{(Color online). Dynamics of the extreme vacuum $\ket{{-}1,3/2,{-}1,{-}3/2}$ in the wake of a quench by Hamiltonian~\eqref{eq:QLM} at $S=3/2$, $\mu/J=\kappa/J=0$, which leads to state-transfer scarring~\cite{Desaules2022a}. (a) The RR~\eqref{eq:RR} shows a cascade of minima related to its various components~\eqref{eq:RRcomponent}. Each minimum corresponds to a maximal overlap with one of the four vacua $\ket{\psi_0^{m_z}}$. The smallest minimum occurs at half the revival period $T{\approx}5.13\pi$ \cite{Desaules2022a}, where the wave function exhibits a very large overlap with the second extreme vacuum $\ket{{-}1,{-}3/2,{-}1,3/2}$. Each DQPT signals a shift in the dominant wave-function overlap within the vacuum manifold. (b) The electric-flux zeros directly connect to the DQPTs signaling a dominance shift in the overlap with the middle vacua $m_z{=}{\pm}1/2$, but other DQPTs do not correspond to zeros in the OP. (c) The minima of the chiral condensate are similar to those of the RR, appearing at roughly the same times.}
	\label{fig:QLM_S3by2} 
\end{figure}

\begin{figure}[t!]
	\centering
	\includegraphics[width=0.48\textwidth]{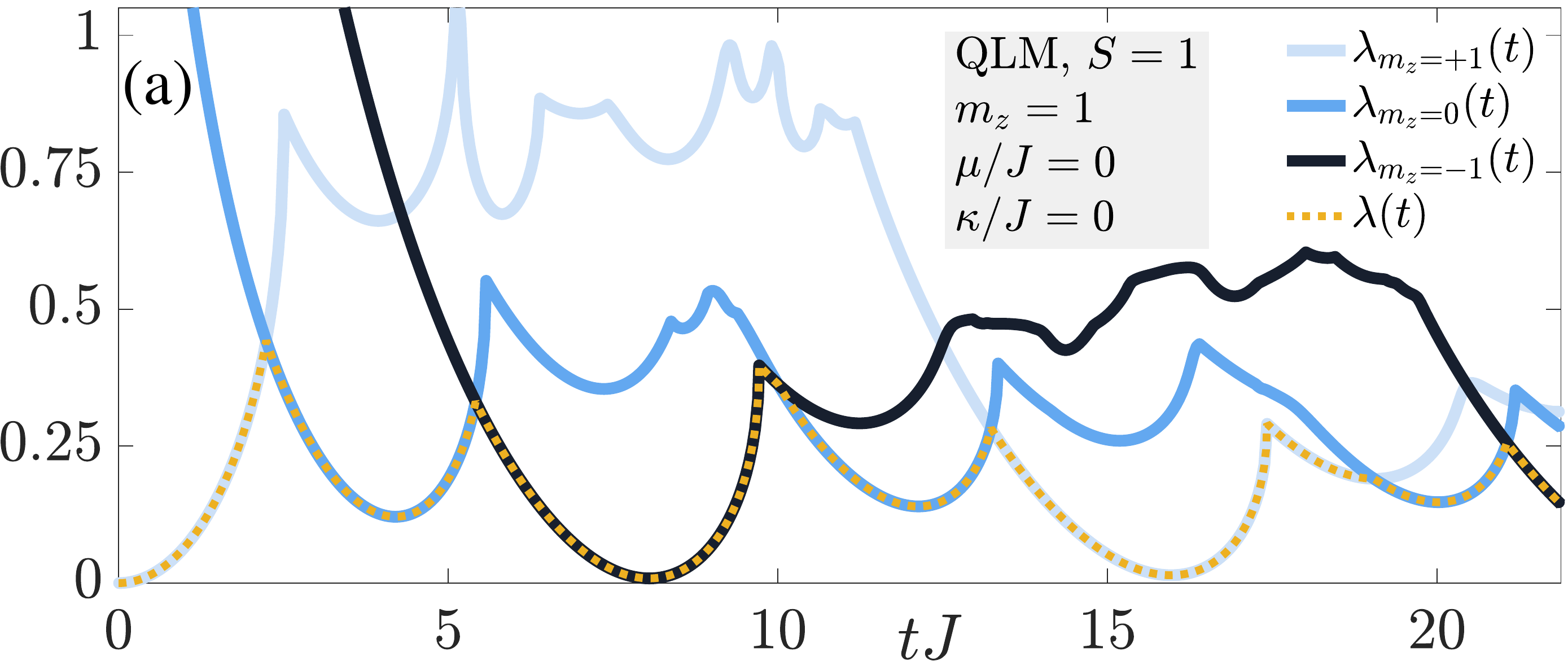}\\
	\vspace{1.1mm}
	\includegraphics[width=0.48\textwidth]{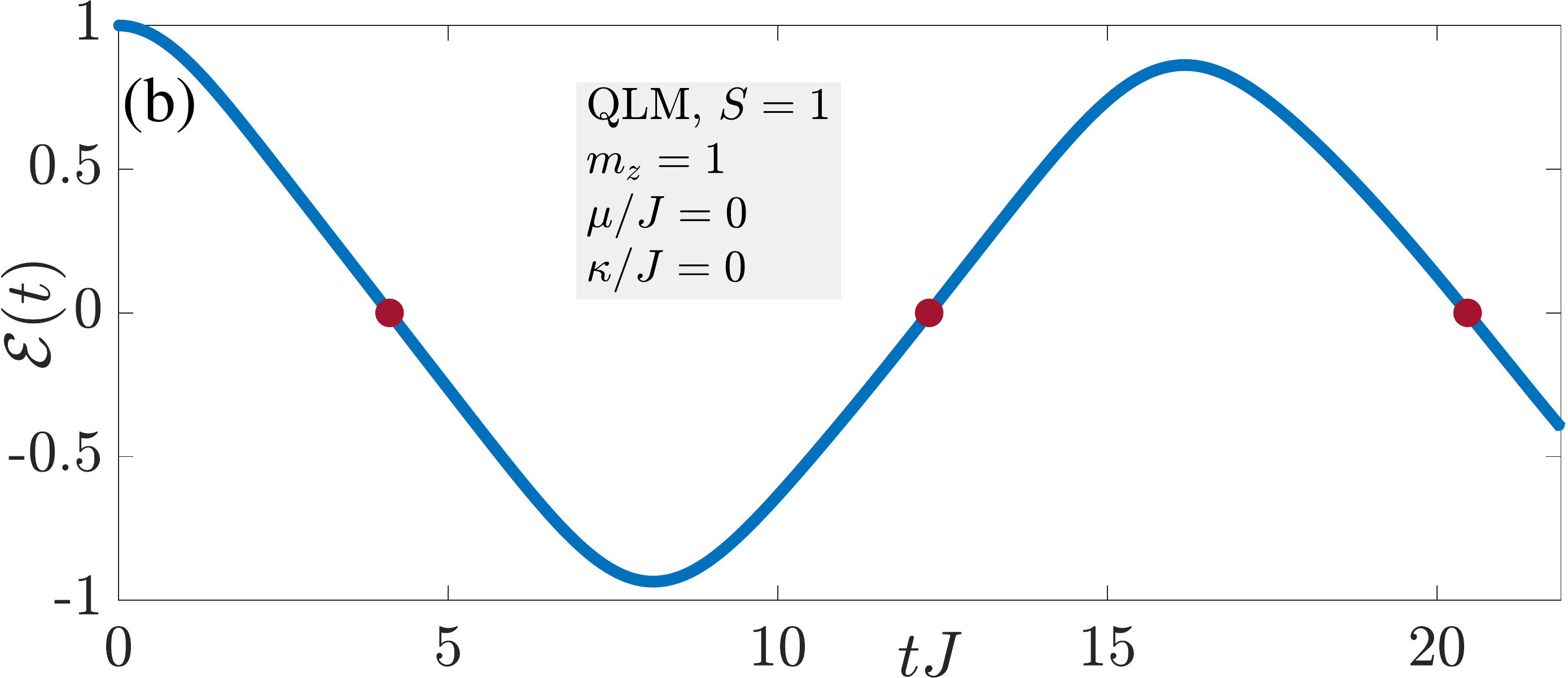}\\
	\vspace{1.1mm}
	\includegraphics[width=0.48\textwidth]{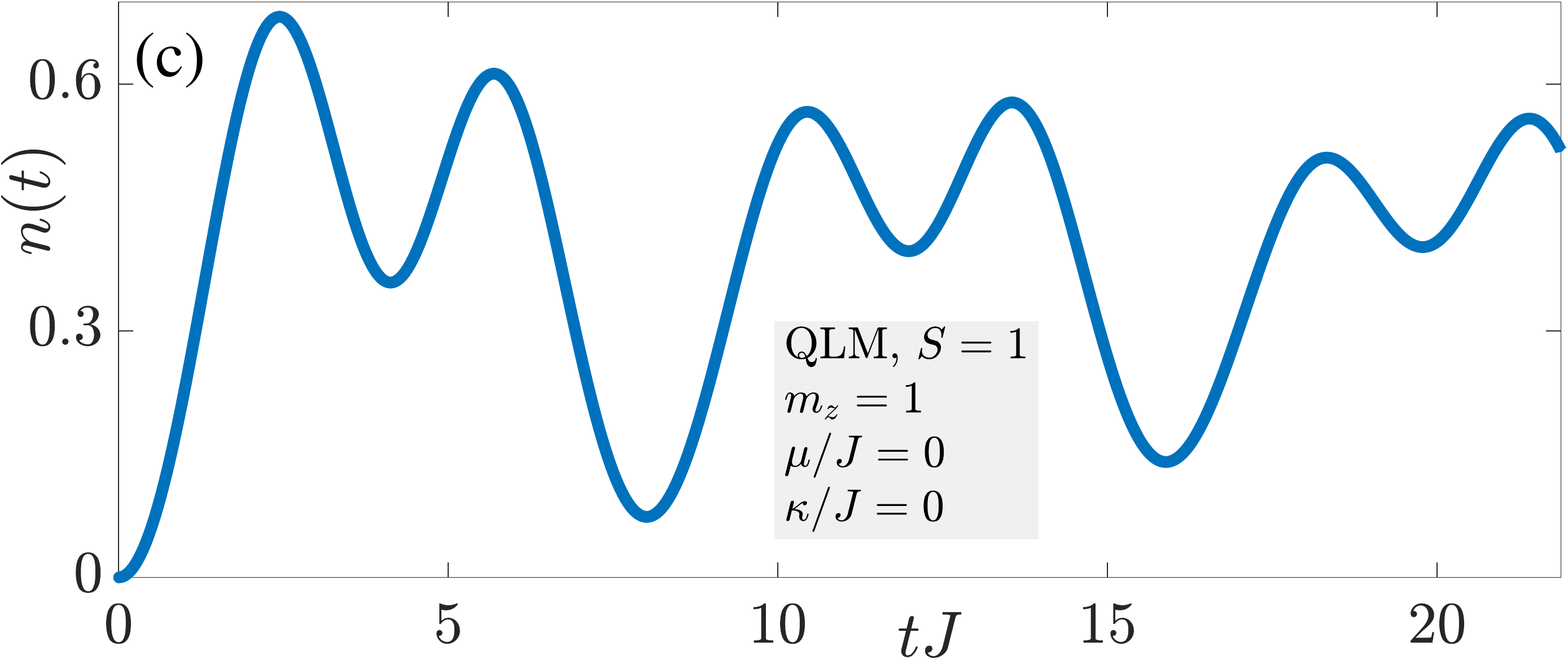}
	\caption{(Color online). Dynamics of the extreme vacuum $\ket{-1,1,-1,0}$ in the wake of a quench by Hamiltonian~\eqref{eq:QLM} at $S=1$, $\mu/J=\kappa/J=0$, which leads to state-transfer scarring~\cite{Desaules2022a}. For integer $S$, the OP zero connects to the MNM of the RR corresponding to the vacuum $m_z{=}0$, and lies at a time between two consecutive DQPTs that signal state transfer to and away from the middle vacuum $m_z{=}0$.}
	\label{fig:QLM_S1}
\end{figure}

We first consider the case of $S{=}3/2$. The corresponding RR~\eqref{eq:RR} and its components~\eqref{eq:RRcomponent} are shown in Fig.~\ref{fig:QLM_S3by2}(a). Focusing on times $t{\lesssim}12.07/J$, we find three DQPTs equally separated in time. The DQPT signaling the state transfer between two vacua forms by the intersection of their corresponding RR-components $\lambda_{m_z}(t)$. The first DQPT indicates state transfer between the extreme vacuum $m_z{=}3/2$ and the intermediate one $m_z{=}1/2$, while the second indicates state transfer between $m_z{=}1/2$ and $m_z{=}{-}1/2$. We find that this second DQPT, which indicates a sign change in the dominant-vacuum flux, occurs at roughly the same time as the first OP zero, marked with a red dot in Fig.~\ref{fig:QLM_S3by2}(b). The two minima in the RR between these DQPTs indicate maximal overlap between the wave function and the intermediate vacua $m_z{=}{\pm}1/2$. The third DQPT signals the dominance of the second extreme vacuum $m_z{=}{-}3/2$ in the wave-function overlap, and at $t{\approx}12.07/J$, we find a minimum in the RR that is very close to zero, indicating maximal overlap with this extreme vacuum. Henceforth, let us call local minima corresponding to extreme (intermediate) vacua as major (minor) local minima, abbreviated as MJM and MNM, respectively. Note how the MJM at $t{\approx}12.07/J$ corresponds to the minimum of the OP, since the second extreme vacuum $m_z{=}{-}3/2$ has the largest negative flux.

For $t{\gtrsim}12.07/J$, the dynamics roughly reverses itself, with two MNM indicating maximal overlaps with the intermediate vacua, and a MJM indicating that the wave function is once again very close to the initial state $m_z{=}3/2$. The state transfer at late times is not as robust as at early times, and so the DQPT signaling transfer between the vacua $m_z{=}{\mp}1/2$ occurs at a slightly different time than the second OP zero: $t{\approx}17.6/J$ and $t{\approx}18.9/J$. This is not surprising, as scarring is expected to deteriorate over time as ergodic dynamics begin to dominate. Indeed, we see a multitude of nonanalyticities in each component $\lambda_{m_z}(t)$ after its first local minimum, which is indicative of complex quantum many-body dynamics. We expect that such nonanalytic behavior will dominate the RR itself at sufficiently long times, eventually destroying scarring and the periodicity of DQPTs.

\begin{figure}[t!]
	\centering
	\includegraphics[width=0.48\textwidth]{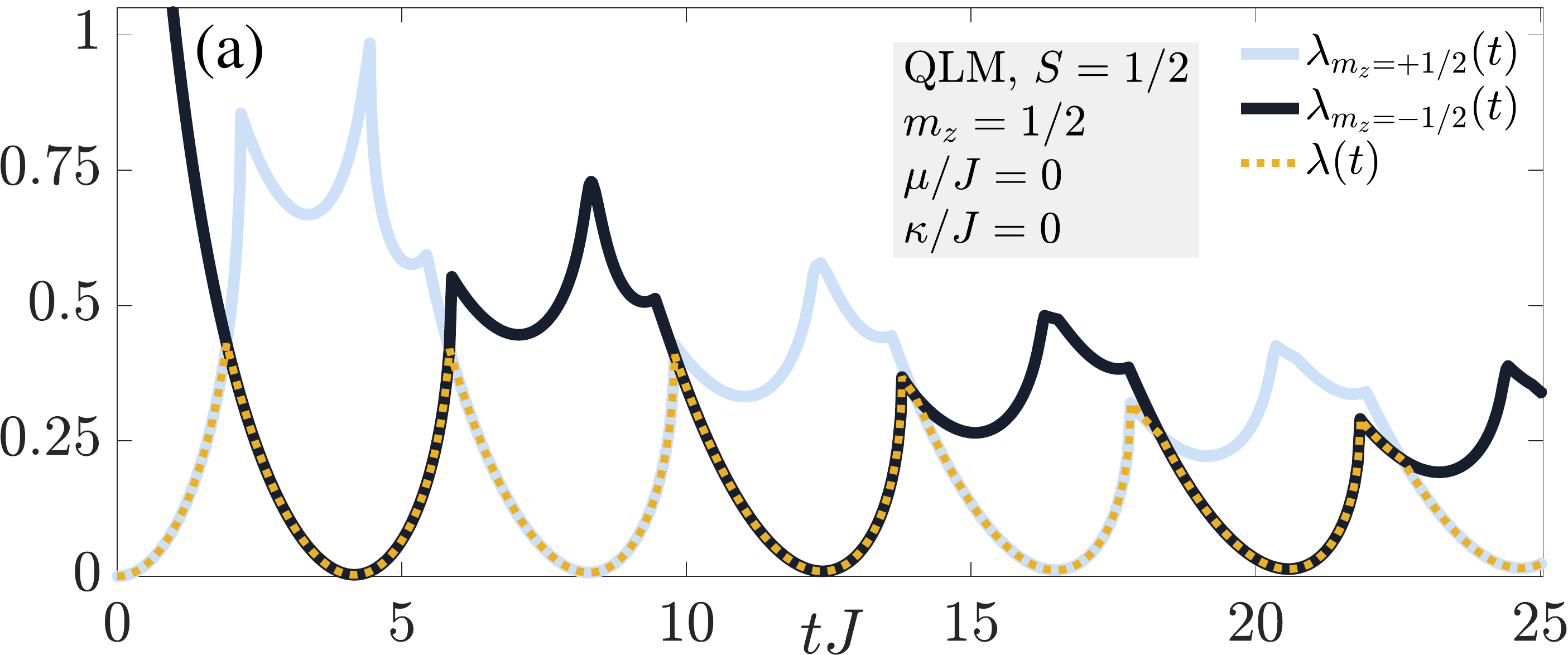}\\
	\vspace{1.1mm}
	\includegraphics[width=0.48\textwidth]{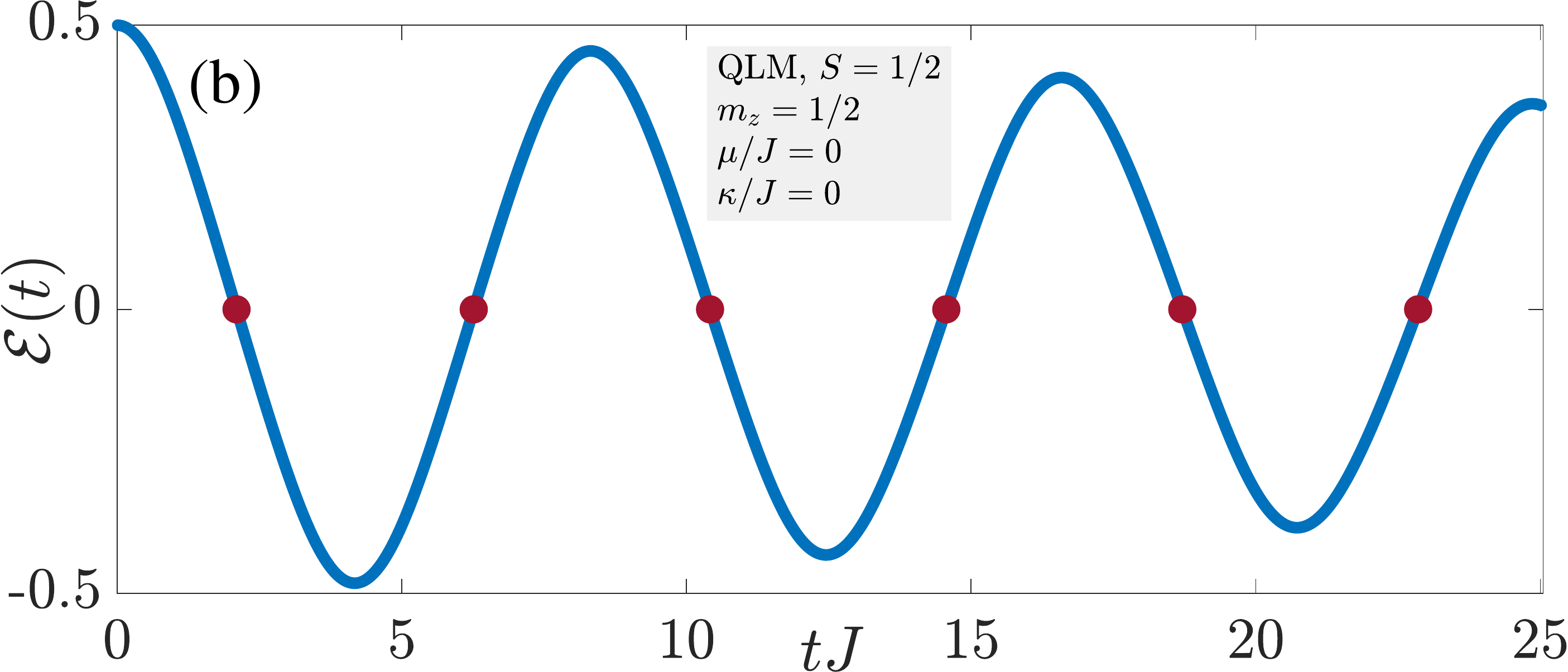}\\
	\vspace{1.1mm}
	\includegraphics[width=0.48\textwidth]{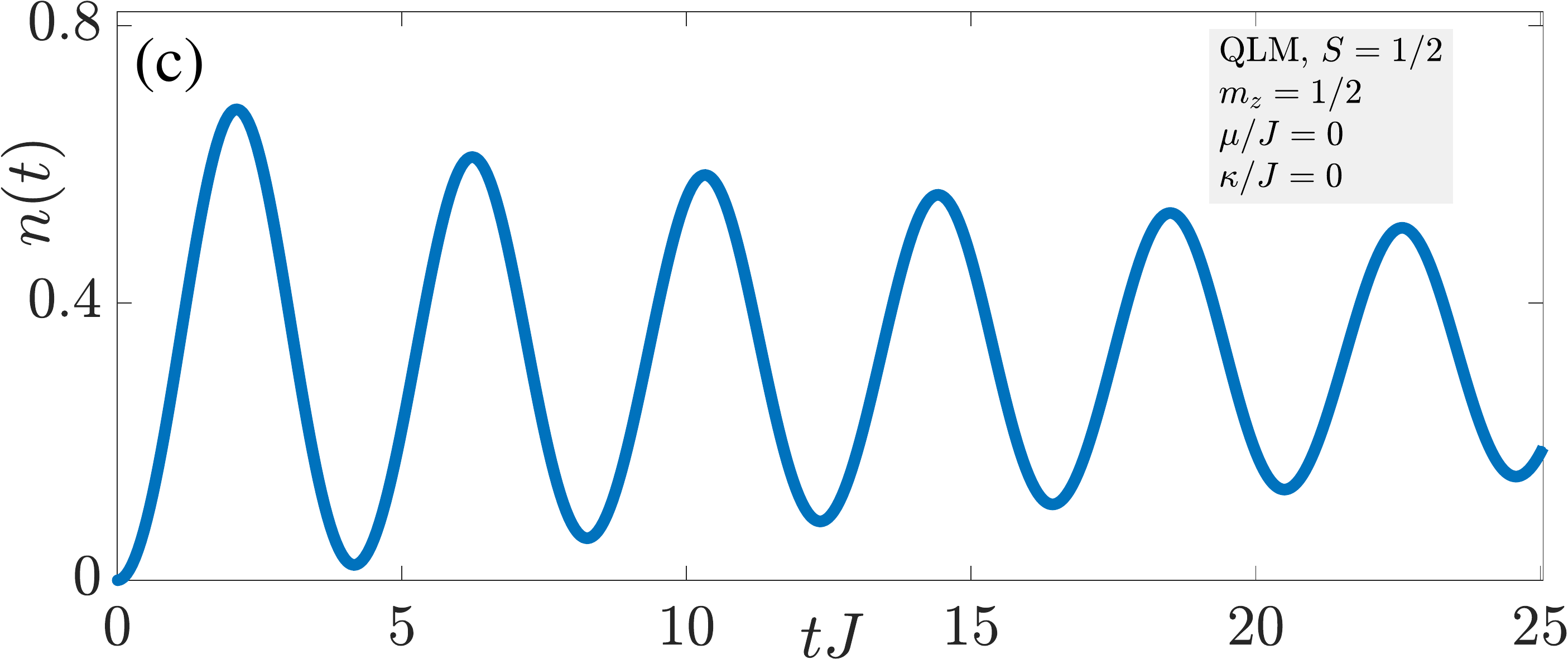}
	\caption{(Color online). Dynamics of the extreme vacuum $\ket{-1,1/2,-1,-1/2}$ in the wake of a quench by Hamiltonian~\eqref{eq:QLM} at $S=1/2$, $\mu/J=\kappa/J=0$, which leads to state-transfer scarring~\cite{Desaules2022a}. For this special case, our general picture reduces to that of the literature, where each DQPT corresponds to an OP zero during evolution times when scarring is robust.}
	\label{fig:QLM_S1by2} 
\end{figure}

An intriguing connection between the chiral condensate~\eqref{eq:CC}, which is not an OP, and the RR~\eqref{eq:RR} can be seen in Fig.~\ref{fig:QLM_S3by2}(c). Every MNM or MJM in the latter is reproduced in the former at the same evolution times, and the relative amplitudes of these local minima are qualitatively similar. In a way, the chiral condensate mirrors the analytic parts in the behavior of the RR, but it does not capture the signatures of DQPTs.

To investigate the effect of the parity of the degenerate manifold, we now repeat the same quench for the spin-$1$ $\mathrm{U}(1)$ QLM, starting in the extreme vacuum $\ket{{-1},1,{-1},{-}1}$, i.e., $m_z{=}S{=}1$. As seen in the fidelity shown in Fig.~\ref{fig:QLM_S1}(a), the phenomenology is the same in that the wave function evolves through large overlaps with the three vacua $m_z{=}0{,}{\pm}1$, where the overlap is largest with the extreme vacua $m_z{=}{\pm}1$. The relation of the RR to the electric flux and chiral condensate, shown in Fig.~\ref{fig:QLM_S1}(b,c), respectively, follows a similar vein as in the case of $S{=}3/2$, except for a fundamental difference in the case of the OP. Whereas for $S{=}3/2$ the DQPT signaling transfer between the vacua $m_z{=}{\pm}1/2$ coincides with an OP zero, for $S{=}1$ no DQPT coincides with an OP zero; see Fig.~\ref{fig:QLM_S1}(b). This is because, for $S{=}1$, the point in time when the OP is zero coincides with the MNM due to maximal wave function overlap with the middle vacuum $m_z{=}0$. This difference only depends on whether $S$ is integer or half-integer, which determines the parity of the $(2S{+}1)$-fold degenerate manifold.

\section{Discussion}
The conclusions drawn from Figs.~\ref{fig:QLM_S3by2} and~\ref{fig:QLM_S1} paint a more general picture of DQPTs, of which previous results resembling two-level dynamics are a special case. In this general picture, a DQPT is associated with a shift in dominance of the RR---equivalently, the wave function---between the different components of the degenerate initial-state manifold. A zero in the OP occurs when the dynamics transfers between different hemispheres of the manifold with opposite OP signs. When the manifold is even-degenerate, the OP zero coincides with the DQPT signaling transfer between the two intermediate vacua with smallest OP absolute value $1/2$. This is what we see in Fig.~\ref{fig:QLM_S3by2} for the spin-$3/2$ $\mathrm{U}(1)$ QLM, and is expected to apply for any half-integer $S$. When the manifold is odd-degenerate, the OP zero does not coincide with a DQPT, but rather occurs half-way in time between two consecutive DQPTs signaling state transfer to and away from the middle vacuum of zero OP. This is what we see in Fig.~\ref{fig:QLM_S3by2} for the spin-$1$ $\mathrm{U}(1)$ QLM, and is expected to hold for any integer $S$. 

The general picture we establish is valid in models with state-transfer scarring. A resemblance to two-level systems, where every DQPT is directly connected to an OP zero and vice versa, is a special case that becomes valid either (i) when the manifold is doubly degenerate, as can be seen in the case of $S{=}1/2$ in Fig.~\ref{fig:QLM_S1by2} where each DQPT is directly connected to an OP zero at short to intermediate times before state transfer slightly deteriorates (see also Ref.~\cite{Huang2019} for small-mass quenches for $S{=}1/2$), or (ii) when the quenches can be perturbatively connected to two-level dynamics, such as in large quenches of quantum Ising models \cite{Halimeh2017,Zunkovic2018,Halimeh2020quasiparticle,Hashizume2022}. Our general picture is also valid for other regularizations of the lattice Schwinger model, and does not apply when state transfer breaks down, e.g., when we quench an extreme vacuum but with a finite mass, or perform a massless quench on an intermediate vacuum (which does not lead to state-transfer dynamics); see Appendix~\ref{app}.

It is also worth noting that our general picture, as well as its special case in previous literature, are only valid for a RR that includes projections on all the states of the initial manifold, as in Eqs.~\eqref{eq:RR_top}. This can be seen by considering, for example, only the component of the initial state itself as the total RR. In such a case we find a plethora of aperiodic DQPTs without any direct connection to the OP; see panels (a) of Figs.~\ref{fig:QLM_S3by2},~\ref{fig:QLM_S1}, and~\ref{fig:QLM_S1by2}. However, we employ the RR defined in Eqs.~\eqref{eq:RR_top} due to its experimental relevance \cite{Jurcevic2017} in addition to its traditional use in the field of DQPTs \cite{Zunkovic2018}.

\bigskip

\begin{acknowledgments}
M.V.D.~acknowledges support from the Research Foundation Flanders (G0E1520N, G0E1820N), and ERC grants QUTE (647905) and ERQUAF (715861). Z.P.~and J.-Y.D.~acknowledge support by EPSRC grant EP/R513258/1 and by the Leverhulme Trust Research Leadership Award RL-2019-015. Z.P. acknowledges support by the Erwin Schr\"odinger International Institute for Mathematics and Physics (ESI). J.C.H.~acknowledges funding from the European Research Council (ERC) under the European Union’s Horizon 2020 research and innovation programm (Grant Agreement no 948141) — ERC Starting Grant SimUcQuam, and by the Deutsche Forschungsgemeinschaft (DFG, German Research Foundation) under Germany's Excellence Strategy -- EXC-2111 -- 390814868.
\end{acknowledgments}

\appendix

\begin{figure}[t!]
	\centering
	\includegraphics[width=0.48\textwidth]{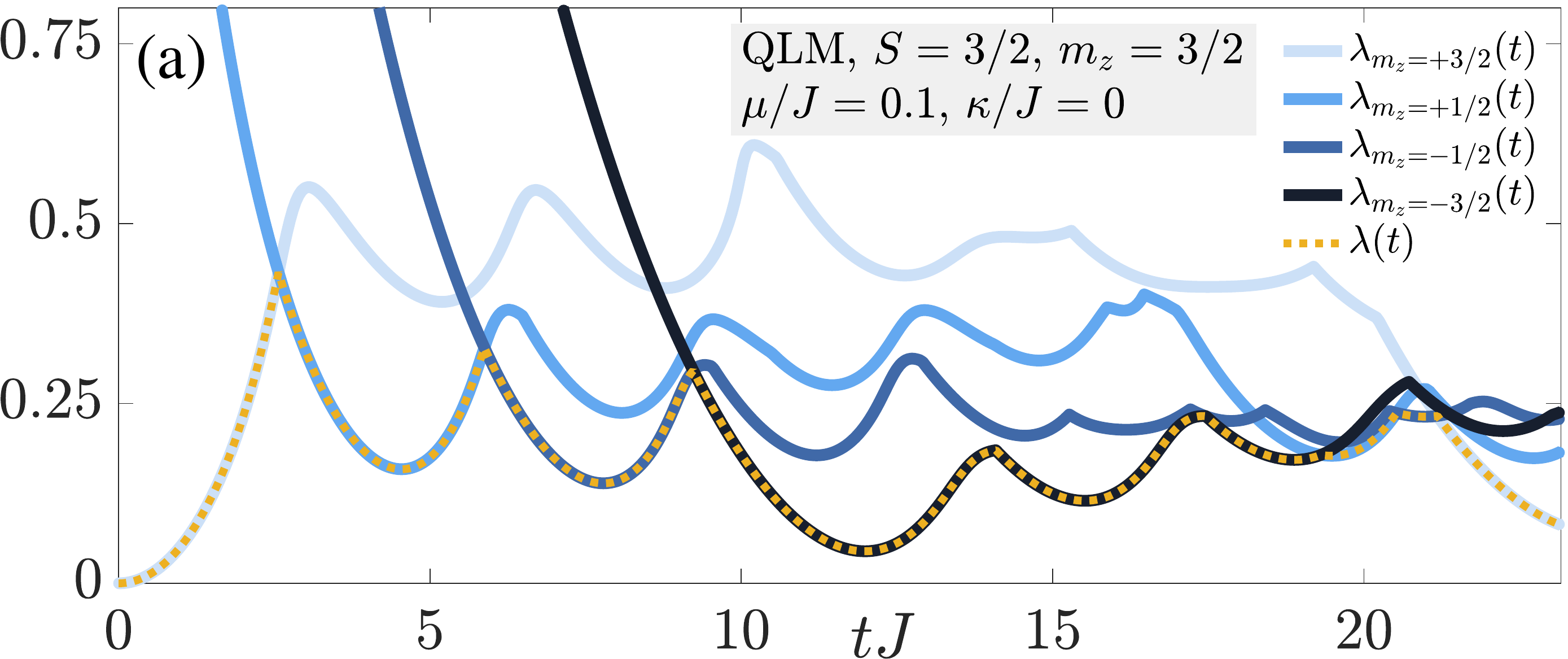}\\
	\vspace{1.1mm}
	\includegraphics[width=0.48\textwidth]{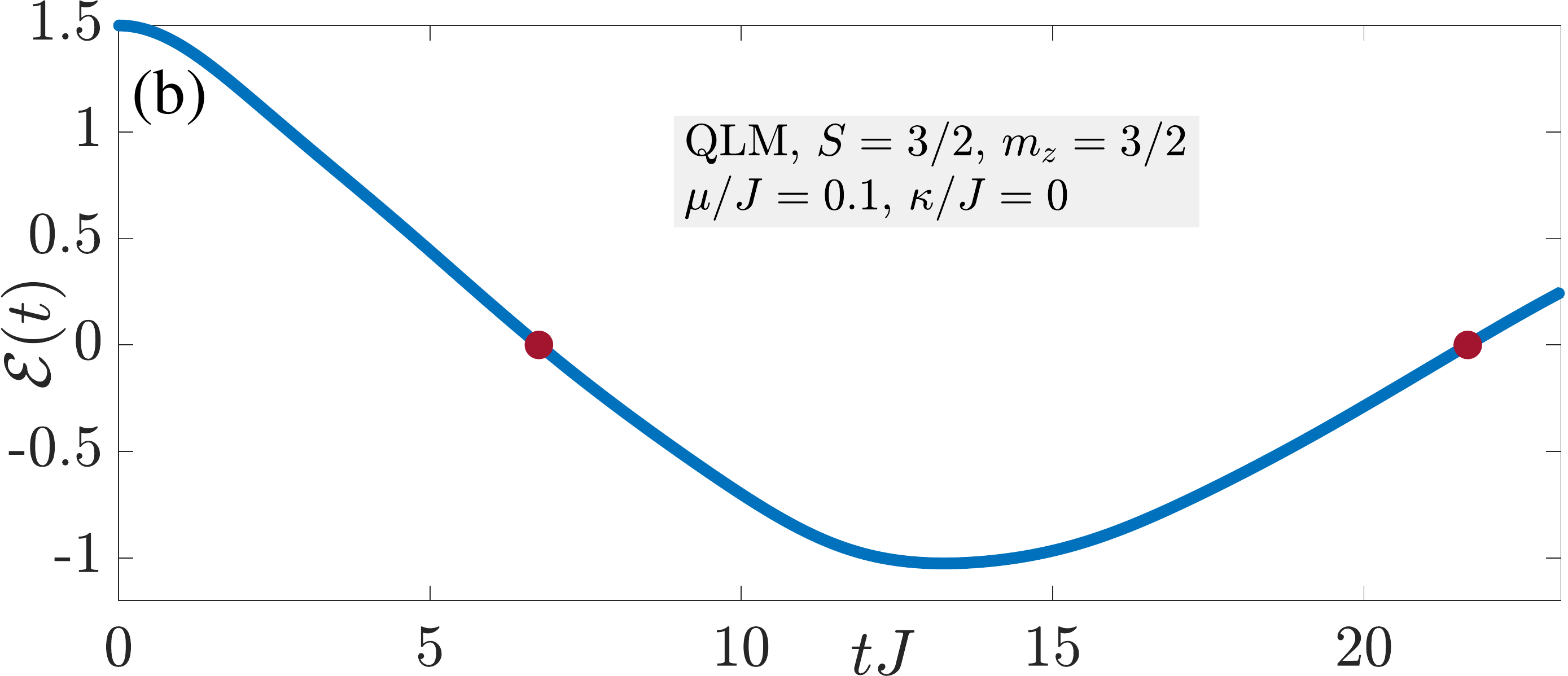}\\
	\vspace{1.1mm}
	\includegraphics[width=0.48\textwidth]{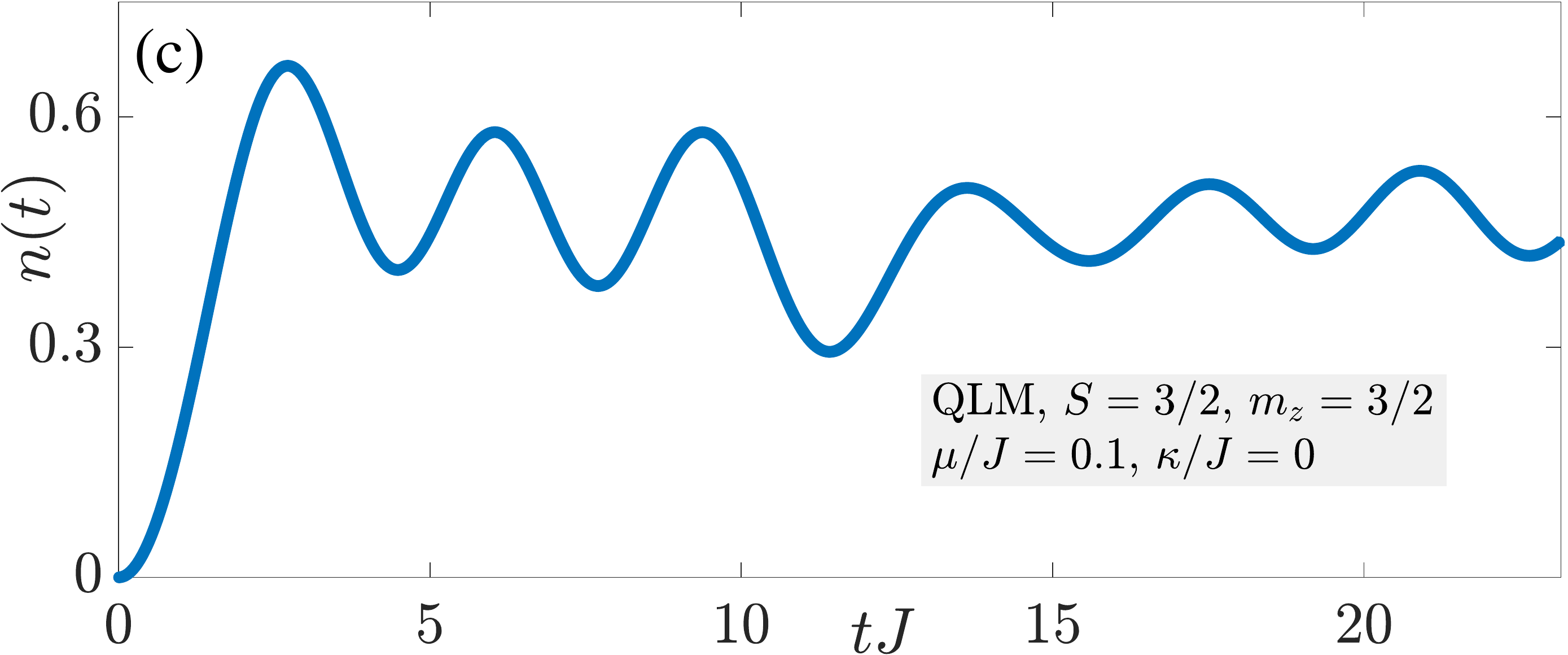}
	\caption{(Color online). Dynamics of the extreme vacuum $\ket{{-}1,3/2,{-}1,{-}3/2}$ in the wake of a quench by Hamiltonian~\eqref{eq:QLM} at $S=3/2$, $\mu/J{=}0.1$ and $\kappa/J{=}0$, which does not lead to state-transfer scarring. This quench is the same as Fig.~\ref{fig:QLM_S3by2} in the main text aside from the value of the mass in the quench Hamiltonian, ($\mu/J{=}0.1$ rather than zero). This quench can be considered perturbatively close to the massless case, where we see the general picture drawn in the main text extends here at early times, but breaks down afterwards.}
	\label{fig:QLM_S3by2_mz3by2_massive} 
\end{figure}

\begin{figure}[t!]
	\centering
	\includegraphics[width=0.48\textwidth]{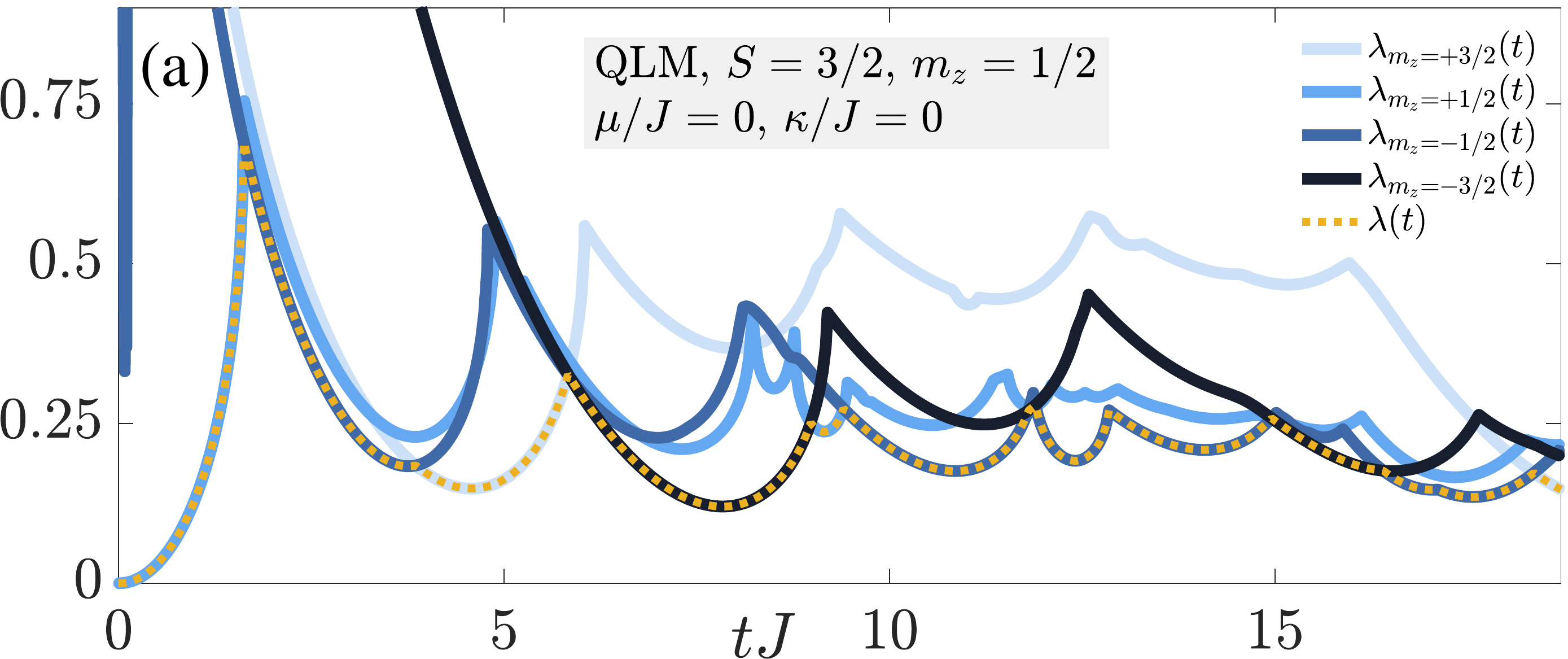}\\
	\vspace{1.1mm}
	\includegraphics[width=0.48\textwidth]{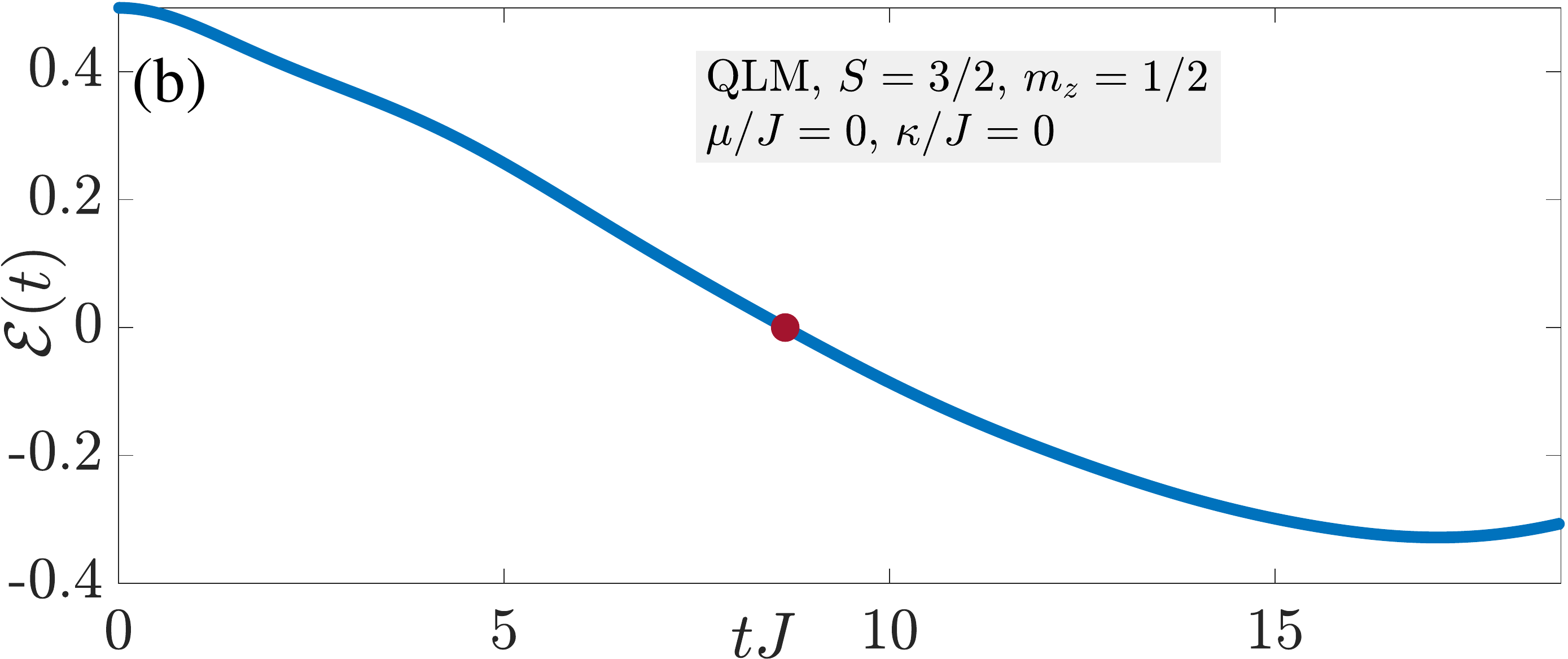}\\
	\vspace{1.1mm}
	\includegraphics[width=0.48\textwidth]{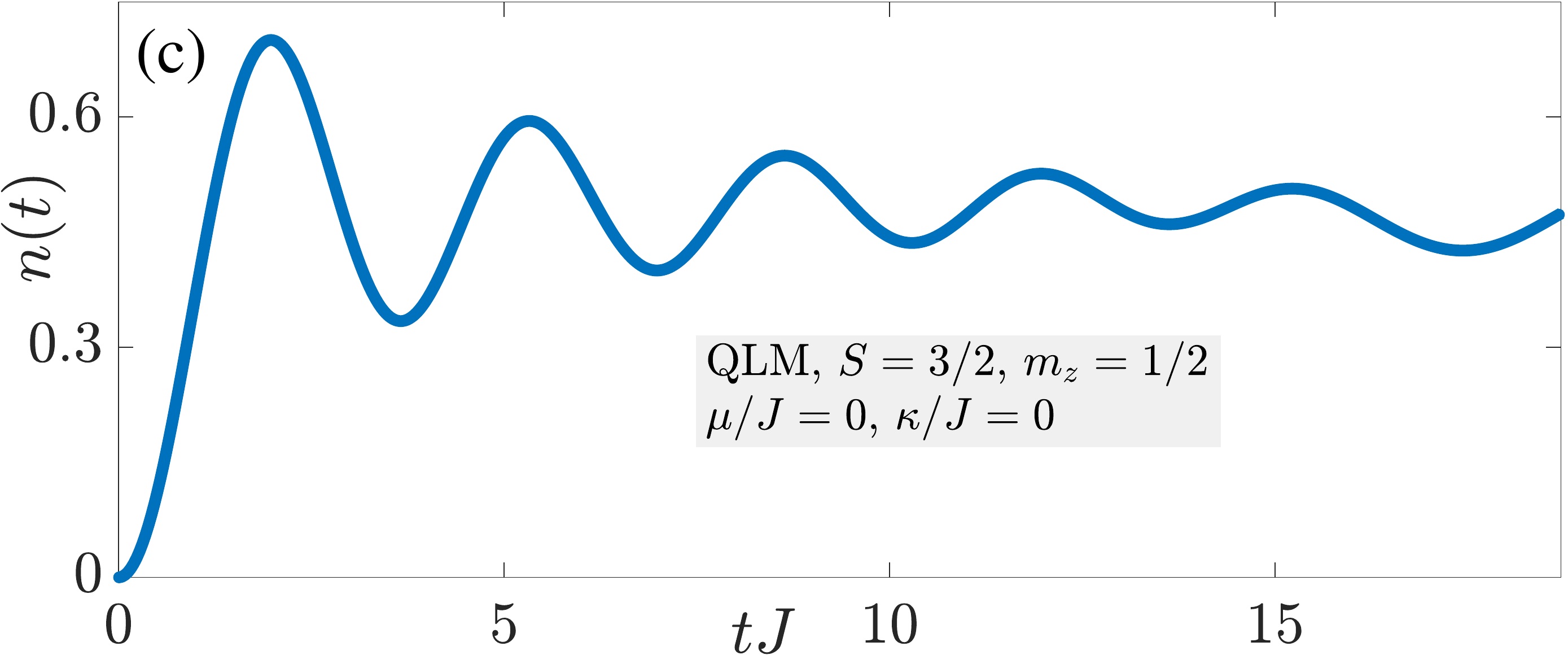}
	\caption{(Color online). Dynamics of the intermediate vacuum $\ket{{-}1,1/2,{-}1,{-}1/2}$ in the wake of a quench by Hamiltonian~\eqref{eq:QLM} at $S=3/2$, $\mu/J{=}0$ and $\kappa/J{=}0$, which does not lead to state-transfer scarring. The only difference between this quench and that of Fig.~\ref{fig:QLM_S3by2} is that the initial state is an intermediate rather than extreme vacuum. Due to the absence of state-transfer dynamics, the general picture we developed in the main text does not apply here.}
	\label{fig:QLM_S3by2_mz1by2} 
\end{figure}

\section{Supplemental results}\label{app}
In this Appendix, we present results for quench protocols that do not lead to state-transfer scarring, leading to the inapplicability of the general picture drawn in the main text. First, we consider a quench of the extreme vacuum in the spin-$3/2$ $\mathrm{U}(1)$ QLM at mass $\mu/J{=}0.1$. Figure~\ref{fig:QLM_S3by2_mz3by2_massive}(a) shows the return rate. At early times, it looks qualitatively similar to that of Fig.~\ref{fig:QLM_S3by2}(a) for the massless quench, but at late times it is qualitatively different, and the periodicity of DQPTs is no longer there. In fact, starting around $t{\approx}19/J$, we see several DQPTs occurring in quick succession. In the general picture of the main text, the first OP zero should coincide with the first DQPT in the case of half-integer $S$, but in this case such a connection is not clearly present; see Fig.~\ref{fig:QLM_S3by2_mz3by2_massive}(b). The dynamics of the chiral condensate, shown in Fig.~\ref{fig:QLM_S3by2_mz3by2_massive}(c), exhibits a connection to the return rate similar to that in the case of the massless quench, where the minima of both quantities roughly coincide in time and relative amplitude.

Next, we consider a massless quench from an intermediate vacuum, where the dynamics becomes significantly different from that of Fig.~\ref{fig:QLM_S3by2}. Indeed, the return rate, shown in Fig.~\ref{fig:QLM_S3by2_mz1by2}(a), exhibits many aperiodic DQPTs, where the OP in the same time window has a single zero that is hard to temporally connect to any of these DQPTs. Even the connection between the chiral condensate, shown in Fig.~\ref{fig:QLM_S3by2_mz1by2}(c), and the return rate is no longer clear.

A different way of regularizing lattice quantum electrodynamics is through the truncated Schwinger model (TSM), given by the Hamiltonian
\begin{align}\nonumber
    \hat{H}_\text{TSM}=\sum_{j=1}^L\bigg[&\frac{J}{2}\big(\hat{\sigma}^-_j\hat{\tau}^+_{j,j+1}\hat{\sigma}^-_{j+1}+\text{H.c.}\big)\\\label{eq:TSM}
    &+\mu\hat{\sigma}^z_j+\frac{\kappa^2}{2}\big(\hat{s}^z_{j,j+1}\big)^2\bigg],
\end{align}
where the only difference with the $\mathrm{U}(1)$ QLM in Eq.~\eqref{eq:QLM} is that the gauge-field operator is given by $\hat{\tau}^+_{j,j+1}$ whose elements are $\big(\hat{\tau}^+_{j,j+1}\big)_{m,n}{=}\delta_{m,n-1}$, instead of $\hat{s}^+_{j,j+1}{/}\sqrt{S(S{+}1)}$. Note that the electric field is still represented by the Pauli operator $\hat{s}^z_{j,j+1}$. Hamiltonian~\eqref{eq:TSM} also hosts a $\mathrm{U}(1)$ gauge symmetry generated by $\hat{G}_j$ in Eq.~\eqref{eq:Gj}, in addition to a global $\mathbb{Z}_2$ symmetry. 

\begin{figure}[t!]
	\centering
	\includegraphics[width=0.48\textwidth]{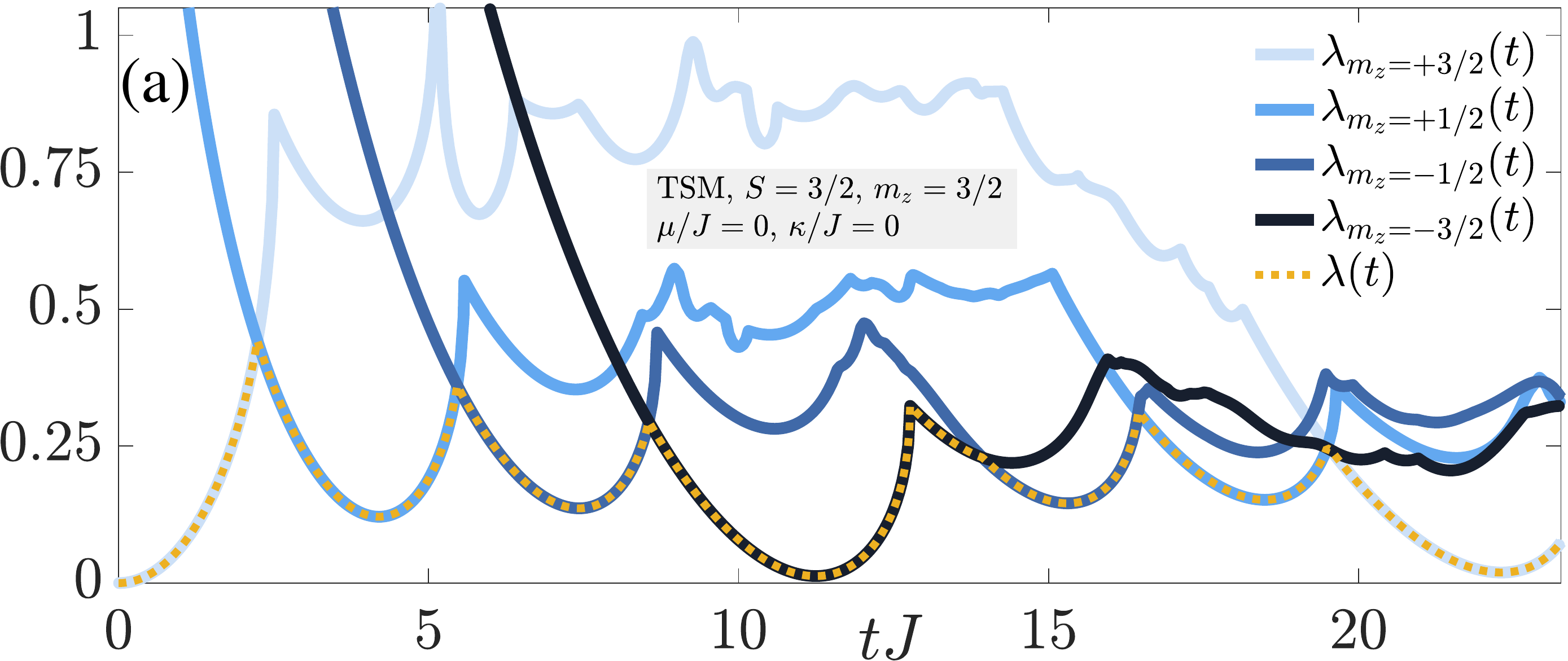}\\
	\vspace{1.1mm}
	\includegraphics[width=0.48\textwidth]{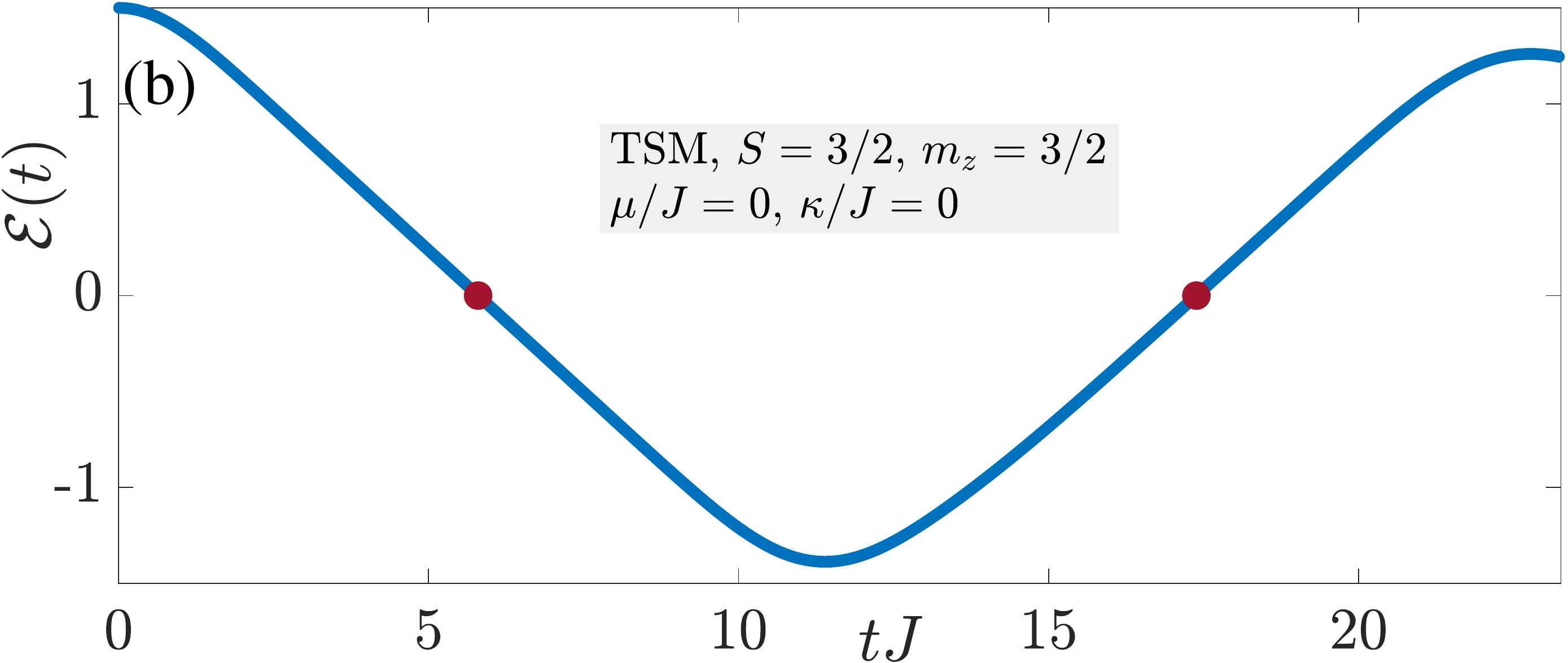}\\
	\vspace{1.1mm}
	\includegraphics[width=0.48\textwidth]{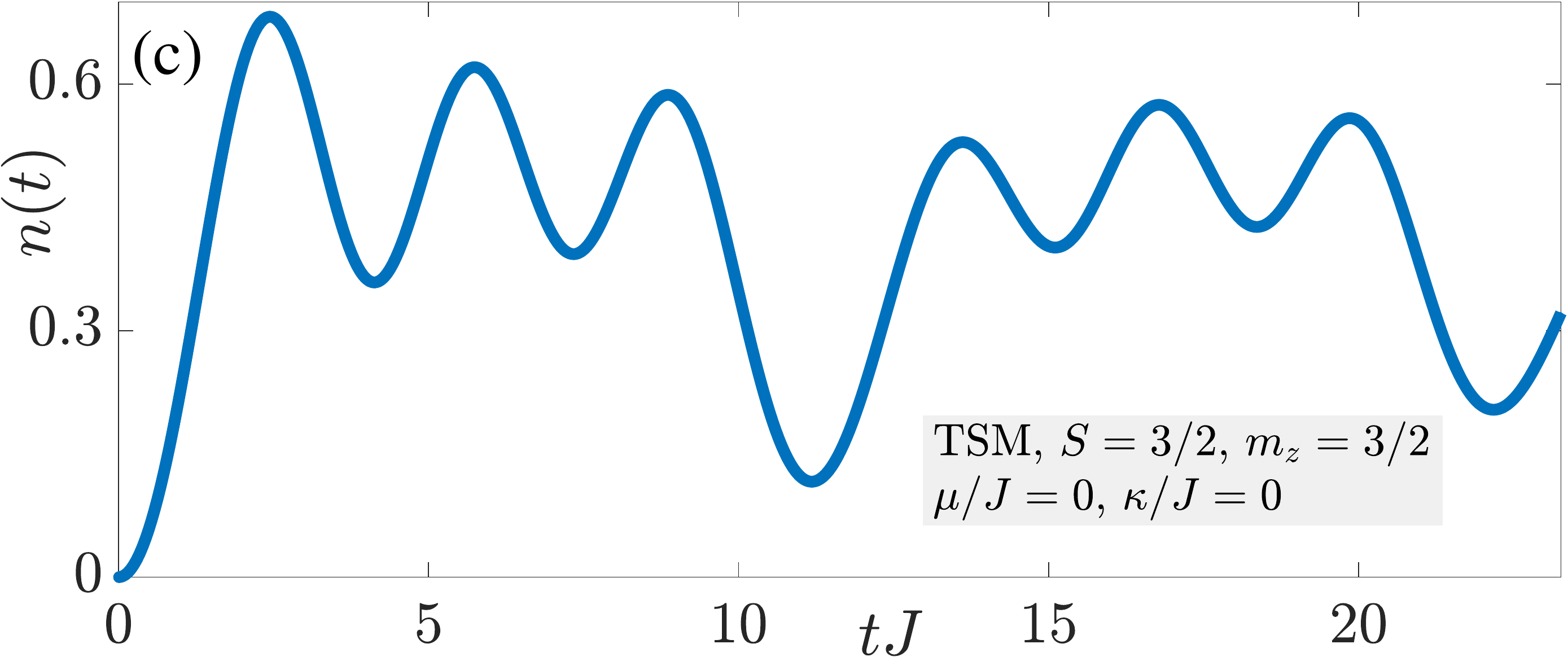}
	\caption{(Color online). Dynamics of the extreme vacuum $\ket{-1,3/2,-1,-3/2}$ in the wake of a quench by Hamiltonian~\eqref{eq:TSM} at $S=3/2$, $\mu/J=\kappa/J=0$, which leads to resonant scarring. The qualitative picture is identical to that of Fig.~\ref{fig:QLM_S3by2} for the spin-$3/2$ $\mathrm{U}(1)$ QLM.}
	\label{fig:TSM_S3by2} 
\end{figure}

The TSM~\eqref{eq:TSM} is equivalent to the QLM~\eqref{eq:QLM} for $S{\leq}1$, up to a trivial rescaling of the tunneling coefficient. As such, we repeat the quench of Fig.~\ref{fig:QLM_S3by2} for the TSM with $S{=}3/2$, where now the time-evolved wave function in Eqs.~\eqref{eq:RR_top} and~\eqref{eq:LocObs} is $\ket{\psi(t)}{=}e^{-i\hat{H}_\text{TSM}t}\ket{\psi_0^S}$. The corresponding quench dynamics is displayed in Fig.~\ref{fig:TSM_S3by2}, where we see qualitatively identical behavior with the corresponding case of the $\mathrm{U}(1)$ QLM. The only difference is that the period is slightly different, $T_\mathrm{TSM}{\approx}11.5\sqrt{S(S{+}1)}$ rather than $T{\approx}5.13\pi S$ for the QLM, confirming results obtained in exact diagonalization for finite system sizes \cite{Desaules2022b}. Again we see the direct one-to-one correspondence in the MNM and MJM of the RR, Fig.~\ref{fig:TSM_S3by2}(a), and those of the chiral condensate, Fig.~\ref{fig:TSM_S3by2}(c). The electric flux, on the other hand, only seems to connect with the MJM of the RR in terms of its minima, and equivalently, the chiral condensate, but does not show anything special at the evolution times where the MNM occur; see Fig.~\ref{fig:TSM_S3by2}(b). As in the case of the QLM, we find that the OP zeros occur at or slightly after the times at which a DQPT arises signaling the shift in wave function-overlap dominance between the two intermediate vacua.

\bibliography{biblio}

\begin{thebibliography}{48}%
\makeatletter
\providecommand \@ifxundefined [1]{%
 \@ifx{#1\undefined}
}%
\providecommand \@ifnum [1]{%
 \ifnum #1\expandafter \@firstoftwo
 \else \expandafter \@secondoftwo
 \fi
}%
\providecommand \@ifx [1]{%
 \ifx #1\expandafter \@firstoftwo
 \else \expandafter \@secondoftwo
 \fi
}%
\providecommand \natexlab [1]{#1}%
\providecommand \enquote  [1]{``#1''}%
\providecommand \bibnamefont  [1]{#1}%
\providecommand \bibfnamefont [1]{#1}%
\providecommand \citenamefont [1]{#1}%
\providecommand \href@noop [0]{\@secondoftwo}%
\providecommand \href [0]{\begingroup \@sanitize@url \@href}%
\providecommand \@href[1]{\@@startlink{#1}\@@href}%
\providecommand \@@href[1]{\endgroup#1\@@endlink}%
\providecommand \@sanitize@url [0]{\catcode `\\12\catcode `\$12\catcode
  `\&12\catcode `\#12\catcode `\^12\catcode `\_12\catcode `\%12\relax}%
\providecommand \@@startlink[1]{}%
\providecommand \@@endlink[0]{}%
\providecommand \url  [0]{\begingroup\@sanitize@url \@url }%
\providecommand \@url [1]{\endgroup\@href {#1}{\urlprefix }}%
\providecommand \urlprefix  [0]{URL }%
\providecommand \Eprint [0]{\href }%
\providecommand \doibase [0]{http://dx.doi.org/}%
\providecommand \selectlanguage [0]{\@gobble}%
\providecommand \bibinfo  [0]{\@secondoftwo}%
\providecommand \bibfield  [0]{\@secondoftwo}%
\providecommand \translation [1]{[#1]}%
\providecommand \BibitemOpen [0]{}%
\providecommand \bibitemStop [0]{}%
\providecommand \bibitemNoStop [0]{.\EOS\space}%
\providecommand \EOS [0]{\spacefactor3000\relax}%
\providecommand \BibitemShut  [1]{\csname bibitem#1\endcsname}%
\let\auto@bib@innerbib\@empty
\bibitem [{\citenamefont {Zvyagin}(2016)}]{Zvyagin2016}%
  \BibitemOpen
  \bibfield  {author} {\bibinfo {author} {\bibfnamefont {A.~A.}\ \bibnamefont
  {Zvyagin}},\ }\bibfield  {title} {\enquote {\bibinfo {title} {Dynamical
  quantum phase transitions ({Review Article})},}\ }\href {\doibase
  10.1063/1.4969869} {\bibfield  {journal} {\bibinfo  {journal} {Low
  Temperature Physics}\ }\textbf {\bibinfo {volume} {42}},\ \bibinfo {pages}
  {971--994} (\bibinfo {year} {2016})}\BibitemShut {NoStop}%
\bibitem [{\citenamefont {Mori}\ \emph {et~al.}(2018)\citenamefont {Mori},
  \citenamefont {Ikeda}, \citenamefont {Kaminishi},\ and\ \citenamefont
  {Ueda}}]{Mori_review}%
  \BibitemOpen
  \bibfield  {author} {\bibinfo {author} {\bibfnamefont {Takashi}\ \bibnamefont
  {Mori}}, \bibinfo {author} {\bibfnamefont {Tatsuhiko~N}\ \bibnamefont
  {Ikeda}}, \bibinfo {author} {\bibfnamefont {Eriko}\ \bibnamefont
  {Kaminishi}}, \ and\ \bibinfo {author} {\bibfnamefont {Masahito}\
  \bibnamefont {Ueda}},\ }\bibfield  {title} {\enquote {\bibinfo {title}
  {Thermalization and prethermalization in isolated quantum systems: a
  theoretical overview},}\ }\href {\doibase 10.1088/1361-6455/aabcdf}
  {\bibfield  {journal} {\bibinfo  {journal} {Journal of Physics B: Atomic,
  Molecular and Optical Physics}\ }\textbf {\bibinfo {volume} {51}},\ \bibinfo
  {pages} {112001} (\bibinfo {year} {2018})}\BibitemShut {NoStop}%
\bibitem [{\citenamefont {Heyl}(2018)}]{Heyl_review}%
  \BibitemOpen
  \bibfield  {author} {\bibinfo {author} {\bibfnamefont {Markus}\ \bibnamefont
  {Heyl}},\ }\bibfield  {title} {\enquote {\bibinfo {title} {Dynamical quantum
  phase transitions: a review},}\ }\href {\doibase 10.1088/1361-6633/aaaf9a}
  {\bibfield  {journal} {\bibinfo  {journal} {Reports on Progress in Physics}\
  }\textbf {\bibinfo {volume} {81}},\ \bibinfo {pages} {054001} (\bibinfo
  {year} {2018})}\BibitemShut {NoStop}%
\bibitem [{\citenamefont {Marino}\ \emph {et~al.}(2022)\citenamefont {Marino},
  \citenamefont {Eckstein}, \citenamefont {Foster},\ and\ \citenamefont
  {Rey}}]{Marino_review}%
  \BibitemOpen
  \bibfield  {author} {\bibinfo {author} {\bibfnamefont {Jamir}\ \bibnamefont
  {Marino}}, \bibinfo {author} {\bibfnamefont {Martin}\ \bibnamefont
  {Eckstein}}, \bibinfo {author} {\bibfnamefont {Matthew}\ \bibnamefont
  {Foster}}, \ and\ \bibinfo {author} {\bibfnamefont {Ana-Maria}\ \bibnamefont
  {Rey}},\ }\bibfield  {title} {\enquote {\bibinfo {title} {Dynamical phase
  transitions in the collisionless pre-thermal states of isolated quantum
  systems: theory and experiments},}\ }\href
  {http://iopscience.iop.org/article/10.1088/1361-6633/ac906c} {\bibfield
  {journal} {\bibinfo  {journal} {Reports on Progress in Physics}\ } (\bibinfo
  {year} {2022})}\BibitemShut {NoStop}%
\bibitem [{\citenamefont {Sciolla}\ and\ \citenamefont
  {Biroli}(2010)}]{Sciolla2010}%
  \BibitemOpen
  \bibfield  {author} {\bibinfo {author} {\bibfnamefont {Bruno}\ \bibnamefont
  {Sciolla}}\ and\ \bibinfo {author} {\bibfnamefont {Giulio}\ \bibnamefont
  {Biroli}},\ }\bibfield  {title} {\enquote {\bibinfo {title} {Quantum quenches
  and off-equilibrium dynamical transition in the infinite-dimensional
  {Bose}-{Hubbard} model},}\ }\href {\doibase 10.1103/PhysRevLett.105.220401}
  {\bibfield  {journal} {\bibinfo  {journal} {Phys. Rev. Lett.}\ }\textbf
  {\bibinfo {volume} {105}},\ \bibinfo {pages} {220401} (\bibinfo {year}
  {2010})}\BibitemShut {NoStop}%
\bibitem [{\citenamefont {Sciolla}\ and\ \citenamefont
  {Biroli}(2011)}]{Sciolla2011}%
  \BibitemOpen
  \bibfield  {author} {\bibinfo {author} {\bibfnamefont {Bruno}\ \bibnamefont
  {Sciolla}}\ and\ \bibinfo {author} {\bibfnamefont {Giulio}\ \bibnamefont
  {Biroli}},\ }\bibfield  {title} {\enquote {\bibinfo {title} {Dynamical
  transitions and quantum quenches in mean-field models},}\ }\href {\doibase
  10.1088/1742-5468/2011/11/p11003} {\bibfield  {journal} {\bibinfo  {journal}
  {Journal of Statistical Mechanics: Theory and Experiment}\ }\textbf {\bibinfo
  {volume} {2011}},\ \bibinfo {pages} {P11003} (\bibinfo {year}
  {2011})}\BibitemShut {NoStop}%
\bibitem [{\citenamefont {Halimeh}\ \emph {et~al.}(2017)\citenamefont
  {Halimeh}, \citenamefont {Zauner-Stauber}, \citenamefont {McCulloch},
  \citenamefont {de~Vega}, \citenamefont {Schollw\"ock},\ and\ \citenamefont
  {Kastner}}]{Halimeh2017a}%
  \BibitemOpen
  \bibfield  {author} {\bibinfo {author} {\bibfnamefont {Jad~C.}\ \bibnamefont
  {Halimeh}}, \bibinfo {author} {\bibfnamefont {Valentin}\ \bibnamefont
  {Zauner-Stauber}}, \bibinfo {author} {\bibfnamefont {Ian~P.}\ \bibnamefont
  {McCulloch}}, \bibinfo {author} {\bibfnamefont {In\'es}\ \bibnamefont
  {de~Vega}}, \bibinfo {author} {\bibfnamefont {Ulrich}\ \bibnamefont
  {Schollw\"ock}}, \ and\ \bibinfo {author} {\bibfnamefont {Michael}\
  \bibnamefont {Kastner}},\ }\bibfield  {title} {\enquote {\bibinfo {title}
  {Prethermalization and persistent order in the absence of a thermal phase
  transition},}\ }\href {\doibase 10.1103/PhysRevB.95.024302} {\bibfield
  {journal} {\bibinfo  {journal} {Phys. Rev. B}\ }\textbf {\bibinfo {volume}
  {95}},\ \bibinfo {pages} {024302} (\bibinfo {year} {2017})}\BibitemShut
  {NoStop}%
\bibitem [{\citenamefont {Heyl}\ \emph {et~al.}(2013)\citenamefont {Heyl},
  \citenamefont {Polkovnikov},\ and\ \citenamefont {Kehrein}}]{Heyl2013}%
  \BibitemOpen
  \bibfield  {author} {\bibinfo {author} {\bibfnamefont {M.}~\bibnamefont
  {Heyl}}, \bibinfo {author} {\bibfnamefont {A.}~\bibnamefont {Polkovnikov}}, \
  and\ \bibinfo {author} {\bibfnamefont {S.}~\bibnamefont {Kehrein}},\
  }\bibfield  {title} {\enquote {\bibinfo {title} {Dynamical quantum phase
  transitions in the transverse-field {Ising} model},}\ }\href {\doibase
  10.1103/PhysRevLett.110.135704} {\bibfield  {journal} {\bibinfo  {journal}
  {Phys. Rev. Lett.}\ }\textbf {\bibinfo {volume} {110}},\ \bibinfo {pages}
  {135704} (\bibinfo {year} {2013})}\BibitemShut {NoStop}%
\bibitem [{\citenamefont {Heyl}(2014)}]{Heyl2014}%
  \BibitemOpen
  \bibfield  {author} {\bibinfo {author} {\bibfnamefont {M.}~\bibnamefont
  {Heyl}},\ }\bibfield  {title} {\enquote {\bibinfo {title} {Dynamical quantum
  phase transitions in systems with broken-symmetry phases},}\ }\href {\doibase
  10.1103/PhysRevLett.113.205701} {\bibfield  {journal} {\bibinfo  {journal}
  {Phys. Rev. Lett.}\ }\textbf {\bibinfo {volume} {113}},\ \bibinfo {pages}
  {205701} (\bibinfo {year} {2014})}\BibitemShut {NoStop}%
\bibitem [{\citenamefont {Halimeh}\ and\ \citenamefont
  {Zauner-Stauber}(2017)}]{Halimeh2017}%
  \BibitemOpen
  \bibfield  {author} {\bibinfo {author} {\bibfnamefont {Jad~C.}\ \bibnamefont
  {Halimeh}}\ and\ \bibinfo {author} {\bibfnamefont {Valentin}\ \bibnamefont
  {Zauner-Stauber}},\ }\bibfield  {title} {\enquote {\bibinfo {title}
  {Dynamical phase diagram of quantum spin chains with long-range
  interactions},}\ }\href {\doibase 10.1103/PhysRevB.96.134427} {\bibfield
  {journal} {\bibinfo  {journal} {Phys. Rev. B}\ }\textbf {\bibinfo {volume}
  {96}},\ \bibinfo {pages} {134427} (\bibinfo {year} {2017})}\BibitemShut
  {NoStop}%
\bibitem [{\citenamefont {Homrighausen}\ \emph {et~al.}(2017)\citenamefont
  {Homrighausen}, \citenamefont {Abeling}, \citenamefont {Zauner-Stauber},\
  and\ \citenamefont {Halimeh}}]{Homrighausen2017}%
  \BibitemOpen
  \bibfield  {author} {\bibinfo {author} {\bibfnamefont {Ingo}\ \bibnamefont
  {Homrighausen}}, \bibinfo {author} {\bibfnamefont {Nils~O.}\ \bibnamefont
  {Abeling}}, \bibinfo {author} {\bibfnamefont {Valentin}\ \bibnamefont
  {Zauner-Stauber}}, \ and\ \bibinfo {author} {\bibfnamefont {Jad~C.}\
  \bibnamefont {Halimeh}},\ }\bibfield  {title} {\enquote {\bibinfo {title}
  {Anomalous dynamical phase in quantum spin chains with long-range
  interactions},}\ }\href {\doibase 10.1103/PhysRevB.96.104436} {\bibfield
  {journal} {\bibinfo  {journal} {Phys. Rev. B}\ }\textbf {\bibinfo {volume}
  {96}},\ \bibinfo {pages} {104436} (\bibinfo {year} {2017})}\BibitemShut
  {NoStop}%
\bibitem [{\citenamefont {\ifmmode \check{Z}\else
  \v{Z}\fi{}unkovi\ifmmode~\check{c}\else \v{c}\fi{}}\ \emph
  {et~al.}(2018)\citenamefont {\ifmmode \check{Z}\else
  \v{Z}\fi{}unkovi\ifmmode~\check{c}\else \v{c}\fi{}}, \citenamefont {Heyl},
  \citenamefont {Knap},\ and\ \citenamefont {Silva}}]{Zunkovic2018}%
  \BibitemOpen
  \bibfield  {author} {\bibinfo {author} {\bibfnamefont {Bojan}\ \bibnamefont
  {\ifmmode \check{Z}\else \v{Z}\fi{}unkovi\ifmmode~\check{c}\else
  \v{c}\fi{}}}, \bibinfo {author} {\bibfnamefont {Markus}\ \bibnamefont
  {Heyl}}, \bibinfo {author} {\bibfnamefont {Michael}\ \bibnamefont {Knap}}, \
  and\ \bibinfo {author} {\bibfnamefont {Alessandro}\ \bibnamefont {Silva}},\
  }\bibfield  {title} {\enquote {\bibinfo {title} {Dynamical quantum phase
  transitions in spin chains with long-range interactions: Merging different
  concepts of nonequilibrium criticality},}\ }\href {\doibase
  10.1103/PhysRevLett.120.130601} {\bibfield  {journal} {\bibinfo  {journal}
  {Phys. Rev. Lett.}\ }\textbf {\bibinfo {volume} {120}},\ \bibinfo {pages}
  {130601} (\bibinfo {year} {2018})}\BibitemShut {NoStop}%
\bibitem [{\citenamefont {Huang}\ \emph {et~al.}(2019)\citenamefont {Huang},
  \citenamefont {Banerjee},\ and\ \citenamefont {Heyl}}]{Huang2019}%
  \BibitemOpen
  \bibfield  {author} {\bibinfo {author} {\bibfnamefont {Yi-Ping}\ \bibnamefont
  {Huang}}, \bibinfo {author} {\bibfnamefont {Debasish}\ \bibnamefont
  {Banerjee}}, \ and\ \bibinfo {author} {\bibfnamefont {Markus}\ \bibnamefont
  {Heyl}},\ }\bibfield  {title} {\enquote {\bibinfo {title} {Dynamical quantum
  phase transitions in {U(1)} quantum link models},}\ }\href {\doibase
  10.1103/PhysRevLett.122.250401} {\bibfield  {journal} {\bibinfo  {journal}
  {Phys. Rev. Lett.}\ }\textbf {\bibinfo {volume} {122}},\ \bibinfo {pages}
  {250401} (\bibinfo {year} {2019})}\BibitemShut {NoStop}%
\bibitem [{\citenamefont {Zache}\ \emph {et~al.}(2019)\citenamefont {Zache},
  \citenamefont {Mueller}, \citenamefont {Schneider}, \citenamefont
  {Jendrzejewski}, \citenamefont {Berges},\ and\ \citenamefont
  {Hauke}}]{Zache2019}%
  \BibitemOpen
  \bibfield  {author} {\bibinfo {author} {\bibfnamefont {T.~V.}\ \bibnamefont
  {Zache}}, \bibinfo {author} {\bibfnamefont {N.}~\bibnamefont {Mueller}},
  \bibinfo {author} {\bibfnamefont {J.~T.}\ \bibnamefont {Schneider}}, \bibinfo
  {author} {\bibfnamefont {F.}~\bibnamefont {Jendrzejewski}}, \bibinfo {author}
  {\bibfnamefont {J.}~\bibnamefont {Berges}}, \ and\ \bibinfo {author}
  {\bibfnamefont {P.}~\bibnamefont {Hauke}},\ }\bibfield  {title} {\enquote
  {\bibinfo {title} {Dynamical topological transitions in the massive
  {Schwinger} model with a $\ensuremath{\theta}$ term},}\ }\href {\doibase
  10.1103/PhysRevLett.122.050403} {\bibfield  {journal} {\bibinfo  {journal}
  {Phys. Rev. Lett.}\ }\textbf {\bibinfo {volume} {122}},\ \bibinfo {pages}
  {050403} (\bibinfo {year} {2019})}\BibitemShut {NoStop}%
\bibitem [{\citenamefont {Halimeh}\ \emph {et~al.}(2020)\citenamefont
  {Halimeh}, \citenamefont {Van~Damme}, \citenamefont {Zauner-Stauber},\ and\
  \citenamefont {Vanderstraeten}}]{Halimeh2020quasiparticle}%
  \BibitemOpen
  \bibfield  {author} {\bibinfo {author} {\bibfnamefont {Jad~C.}\ \bibnamefont
  {Halimeh}}, \bibinfo {author} {\bibfnamefont {Maarten}\ \bibnamefont
  {Van~Damme}}, \bibinfo {author} {\bibfnamefont {Valentin}\ \bibnamefont
  {Zauner-Stauber}}, \ and\ \bibinfo {author} {\bibfnamefont {Laurens}\
  \bibnamefont {Vanderstraeten}},\ }\bibfield  {title} {\enquote {\bibinfo
  {title} {Quasiparticle origin of dynamical quantum phase transitions},}\
  }\href {\doibase 10.1103/PhysRevResearch.2.033111} {\bibfield  {journal}
  {\bibinfo  {journal} {Phys. Rev. Research}\ }\textbf {\bibinfo {volume}
  {2}},\ \bibinfo {pages} {033111} (\bibinfo {year} {2020})}\BibitemShut
  {NoStop}%
\bibitem [{\citenamefont {Hashizume}\ \emph {et~al.}(2022)\citenamefont
  {Hashizume}, \citenamefont {McCulloch},\ and\ \citenamefont
  {Halimeh}}]{Hashizume2022}%
  \BibitemOpen
  \bibfield  {author} {\bibinfo {author} {\bibfnamefont {Tomohiro}\
  \bibnamefont {Hashizume}}, \bibinfo {author} {\bibfnamefont {Ian~P.}\
  \bibnamefont {McCulloch}}, \ and\ \bibinfo {author} {\bibfnamefont {Jad~C.}\
  \bibnamefont {Halimeh}},\ }\bibfield  {title} {\enquote {\bibinfo {title}
  {Dynamical phase transitions in the two-dimensional transverse-field {Ising}
  model},}\ }\href {\doibase 10.1103/PhysRevResearch.4.013250} {\bibfield
  {journal} {\bibinfo  {journal} {Phys. Rev. Research}\ }\textbf {\bibinfo
  {volume} {4}},\ \bibinfo {pages} {013250} (\bibinfo {year}
  {2022})}\BibitemShut {NoStop}%
\bibitem [{\citenamefont {Andraschko}\ and\ \citenamefont
  {Sirker}(2014)}]{Sirker2014}%
  \BibitemOpen
  \bibfield  {author} {\bibinfo {author} {\bibfnamefont {F.}~\bibnamefont
  {Andraschko}}\ and\ \bibinfo {author} {\bibfnamefont {J.}~\bibnamefont
  {Sirker}},\ }\bibfield  {title} {\enquote {\bibinfo {title} {Dynamical
  quantum phase transitions and the {Loschmidt} echo: A transfer matrix
  approach},}\ }\href {\doibase 10.1103/PhysRevB.89.125120} {\bibfield
  {journal} {\bibinfo  {journal} {Phys. Rev. B}\ }\textbf {\bibinfo {volume}
  {89}},\ \bibinfo {pages} {125120} (\bibinfo {year} {2014})}\BibitemShut
  {NoStop}%
\bibitem [{\citenamefont {Vajna}\ and\ \citenamefont
  {D\'ora}(2014)}]{Vajna2014}%
  \BibitemOpen
  \bibfield  {author} {\bibinfo {author} {\bibfnamefont {Szabolcs}\
  \bibnamefont {Vajna}}\ and\ \bibinfo {author} {\bibfnamefont {Bal\'azs}\
  \bibnamefont {D\'ora}},\ }\bibfield  {title} {\enquote {\bibinfo {title}
  {Disentangling dynamical phase transitions from equilibrium phase
  transitions},}\ }\href {\doibase 10.1103/PhysRevB.89.161105} {\bibfield
  {journal} {\bibinfo  {journal} {Phys. Rev. B}\ }\textbf {\bibinfo {volume}
  {89}},\ \bibinfo {pages} {161105} (\bibinfo {year} {2014})}\BibitemShut
  {NoStop}%
\bibitem [{\citenamefont {Karrasch}\ and\ \citenamefont
  {Schuricht}(2013)}]{Karrasch2013}%
  \BibitemOpen
  \bibfield  {author} {\bibinfo {author} {\bibfnamefont {C.}~\bibnamefont
  {Karrasch}}\ and\ \bibinfo {author} {\bibfnamefont {D.}~\bibnamefont
  {Schuricht}},\ }\bibfield  {title} {\enquote {\bibinfo {title} {Dynamical
  phase transitions after quenches in nonintegrable models},}\ }\href {\doibase
  10.1103/PhysRevB.87.195104} {\bibfield  {journal} {\bibinfo  {journal} {Phys.
  Rev. B}\ }\textbf {\bibinfo {volume} {87}},\ \bibinfo {pages} {195104}
  (\bibinfo {year} {2013})}\BibitemShut {NoStop}%
\bibitem [{\citenamefont {Corps}\ and\ \citenamefont
  {Rela\~no}(2022)}]{Corps2022}%
  \BibitemOpen
  \bibfield  {author} {\bibinfo {author} {\bibfnamefont {\'Angel~L.}\
  \bibnamefont {Corps}}\ and\ \bibinfo {author} {\bibfnamefont {Armando}\
  \bibnamefont {Rela\~no}},\ }\bibfield  {title} {\enquote {\bibinfo {title}
  {Dynamical and excited-state quantum phase transitions in collective
  systems},}\ }\href {\doibase 10.1103/PhysRevB.106.024311} {\bibfield
  {journal} {\bibinfo  {journal} {Phys. Rev. B}\ }\textbf {\bibinfo {volume}
  {106}},\ \bibinfo {pages} {024311} (\bibinfo {year} {2022})}\BibitemShut
  {NoStop}%
\bibitem [{\citenamefont {Zakrzewski}(2022)}]{Zakrzewski2022}%
  \BibitemOpen
  \bibfield  {author} {\bibinfo {author} {\bibfnamefont {Jakub}\ \bibnamefont
  {Zakrzewski}},\ }\bibfield  {title} {\enquote {\bibinfo {title} {Dynamical
  quantum phase transitions from quantum optics perspective},}\ }\href@noop {}
  {\bibfield  {journal} {\bibinfo  {journal} {arXiv e-prints}\ } (\bibinfo
  {year} {2022})},\ \Eprint {http://arxiv.org/abs/2204.09454} {arXiv:2204.09454
  [quant-ph]} \BibitemShut {NoStop}%
\bibitem [{\citenamefont {Serbyn}\ \emph {et~al.}(2021)\citenamefont {Serbyn},
  \citenamefont {Abanin},\ and\ \citenamefont {Papi{\'c}}}]{Serbyn2021}%
  \BibitemOpen
  \bibfield  {author} {\bibinfo {author} {\bibfnamefont {Maksym}\ \bibnamefont
  {Serbyn}}, \bibinfo {author} {\bibfnamefont {Dmitry~A.}\ \bibnamefont
  {Abanin}}, \ and\ \bibinfo {author} {\bibfnamefont {Zlatko}\ \bibnamefont
  {Papi{\'c}}},\ }\bibfield  {title} {\enquote {\bibinfo {title} {Quantum
  many-body scars and weak breaking of ergodicity},}\ }\href {\doibase
  10.1038/s41567-021-01230-2} {\bibfield  {journal} {\bibinfo  {journal}
  {Nature Physics}\ }\textbf {\bibinfo {volume} {17}},\ \bibinfo {pages}
  {675--685} (\bibinfo {year} {2021})}\BibitemShut {NoStop}%
\bibitem [{\citenamefont {Moudgalya}\ \emph {et~al.}(2022)\citenamefont
  {Moudgalya}, \citenamefont {Bernevig},\ and\ \citenamefont
  {Regnault}}]{MoudgalyaReview}%
  \BibitemOpen
  \bibfield  {author} {\bibinfo {author} {\bibfnamefont {Sanjay}\ \bibnamefont
  {Moudgalya}}, \bibinfo {author} {\bibfnamefont {B~Andrei}\ \bibnamefont
  {Bernevig}}, \ and\ \bibinfo {author} {\bibfnamefont {Nicolas}\ \bibnamefont
  {Regnault}},\ }\bibfield  {title} {\enquote {\bibinfo {title} {Quantum
  many-body scars and {Hilbert} space fragmentation: a review of exact
  results},}\ }\href {\doibase 10.1088/1361-6633/ac73a0} {\bibfield  {journal}
  {\bibinfo  {journal} {Reports on Progress in Physics}\ }\textbf {\bibinfo
  {volume} {85}},\ \bibinfo {pages} {086501} (\bibinfo {year}
  {2022})}\BibitemShut {NoStop}%
\bibitem [{\citenamefont {Chandran}\ \emph {et~al.}(2022)\citenamefont
  {Chandran}, \citenamefont {Iadecola}, \citenamefont {Khemani},\ and\
  \citenamefont {Moessner}}]{ChandranReview}%
  \BibitemOpen
  \bibfield  {author} {\bibinfo {author} {\bibfnamefont {Anushya}\ \bibnamefont
  {Chandran}}, \bibinfo {author} {\bibfnamefont {Thomas}\ \bibnamefont
  {Iadecola}}, \bibinfo {author} {\bibfnamefont {Vedika}\ \bibnamefont
  {Khemani}}, \ and\ \bibinfo {author} {\bibfnamefont {Roderich}\ \bibnamefont
  {Moessner}},\ }\href {\doibase 10.48550/ARXIV.2206.11528} {\enquote {\bibinfo
  {title} {Quantum many-body scars: A quasiparticle perspective},}\ } (\bibinfo
  {year} {2022})\BibitemShut {NoStop}%
\bibitem [{\citenamefont {Schwinger}(1962)}]{Schwinger1962}%
  \BibitemOpen
  \bibfield  {author} {\bibinfo {author} {\bibfnamefont {Julian}\ \bibnamefont
  {Schwinger}},\ }\bibfield  {title} {\enquote {\bibinfo {title} {Gauge
  invariance and mass. {II}},}\ }\href {\doibase 10.1103/PhysRev.128.2425}
  {\bibfield  {journal} {\bibinfo  {journal} {Phys. Rev.}\ }\textbf {\bibinfo
  {volume} {128}},\ \bibinfo {pages} {2425--2429} (\bibinfo {year}
  {1962})}\BibitemShut {NoStop}%
\bibitem [{\citenamefont {Chandrasekharan}\ and\ \citenamefont
  {Wiese}(1997)}]{Chandrasekharan1997}%
  \BibitemOpen
  \bibfield  {author} {\bibinfo {author} {\bibfnamefont {S}~\bibnamefont
  {Chandrasekharan}}\ and\ \bibinfo {author} {\bibfnamefont {U.-J}\
  \bibnamefont {Wiese}},\ }\bibfield  {title} {\enquote {\bibinfo {title}
  {Quantum link models: A discrete approach to gauge theories},}\ }\href
  {\doibase https://doi.org/10.1016/S0550-3213(97)80041-7} {\bibfield
  {journal} {\bibinfo  {journal} {Nuclear Physics B}\ }\textbf {\bibinfo
  {volume} {492}},\ \bibinfo {pages} {455 -- 471} (\bibinfo {year}
  {1997})}\BibitemShut {NoStop}%
\bibitem [{\citenamefont {Wiese}(2013)}]{Wiese_review}%
  \BibitemOpen
  \bibfield  {author} {\bibinfo {author} {\bibfnamefont {U.-J.}\ \bibnamefont
  {Wiese}},\ }\bibfield  {title} {\enquote {\bibinfo {title} {Ultracold quantum
  gases and lattice systems: quantum simulation of lattice gauge theories},}\
  }\href {\doibase 10.1002/andp.201300104} {\bibfield  {journal} {\bibinfo
  {journal} {Annalen der Physik}\ }\textbf {\bibinfo {volume} {525}},\ \bibinfo
  {pages} {777--796} (\bibinfo {year} {2013})}\BibitemShut {NoStop}%
\bibitem [{\citenamefont {Surace}\ \emph {et~al.}(2020)\citenamefont {Surace},
  \citenamefont {Mazza}, \citenamefont {Giudici}, \citenamefont {Lerose},
  \citenamefont {Gambassi},\ and\ \citenamefont {Dalmonte}}]{Surace2020}%
  \BibitemOpen
  \bibfield  {author} {\bibinfo {author} {\bibfnamefont {Federica~M.}\
  \bibnamefont {Surace}}, \bibinfo {author} {\bibfnamefont {Paolo~P.}\
  \bibnamefont {Mazza}}, \bibinfo {author} {\bibfnamefont {Giuliano}\
  \bibnamefont {Giudici}}, \bibinfo {author} {\bibfnamefont {Alessio}\
  \bibnamefont {Lerose}}, \bibinfo {author} {\bibfnamefont {Andrea}\
  \bibnamefont {Gambassi}}, \ and\ \bibinfo {author} {\bibfnamefont {Marcello}\
  \bibnamefont {Dalmonte}},\ }\bibfield  {title} {\enquote {\bibinfo {title}
  {Lattice gauge theories and string dynamics in rydberg atom quantum
  simulators},}\ }\href {\doibase 10.1103/PhysRevX.10.021041} {\bibfield
  {journal} {\bibinfo  {journal} {Phys. Rev. X}\ }\textbf {\bibinfo {volume}
  {10}},\ \bibinfo {pages} {021041} (\bibinfo {year} {2020})}\BibitemShut
  {NoStop}%
\bibitem [{\citenamefont {{Desaules}}\ \emph
  {et~al.}(2022{\natexlab{a}})\citenamefont {{Desaules}}, \citenamefont
  {{Banerjee}}, \citenamefont {{Hudomal}}, \citenamefont {{Papi{\'c}}},
  \citenamefont {{Sen}},\ and\ \citenamefont {{Halimeh}}}]{Desaules2022a}%
  \BibitemOpen
  \bibfield  {author} {\bibinfo {author} {\bibfnamefont {Jean-Yves}\
  \bibnamefont {{Desaules}}}, \bibinfo {author} {\bibfnamefont {Debasish}\
  \bibnamefont {{Banerjee}}}, \bibinfo {author} {\bibfnamefont {Ana}\
  \bibnamefont {{Hudomal}}}, \bibinfo {author} {\bibfnamefont {Zlatko}\
  \bibnamefont {{Papi{\'c}}}}, \bibinfo {author} {\bibfnamefont {Arnab}\
  \bibnamefont {{Sen}}}, \ and\ \bibinfo {author} {\bibfnamefont {Jad~C.}\
  \bibnamefont {{Halimeh}}},\ }\bibfield  {title} {\enquote {\bibinfo {title}
  {Weak ergodicity breaking in the {Schwinger} model},}\ }\href@noop {}
  {\bibfield  {journal} {\bibinfo  {journal} {arXiv e-prints}\ } (\bibinfo
  {year} {2022}{\natexlab{a}})},\ \Eprint {http://arxiv.org/abs/2203.08830}
  {arXiv:2203.08830 [cond-mat.str-el]} \BibitemShut {NoStop}%
\bibitem [{\citenamefont {Kasper}\ \emph {et~al.}(2017)\citenamefont {Kasper},
  \citenamefont {Hebenstreit}, \citenamefont {Jendrzejewski}, \citenamefont
  {Oberthaler},\ and\ \citenamefont {Berges}}]{Kasper2017}%
  \BibitemOpen
  \bibfield  {author} {\bibinfo {author} {\bibfnamefont {V.}~\bibnamefont
  {Kasper}}, \bibinfo {author} {\bibfnamefont {F.}~\bibnamefont {Hebenstreit}},
  \bibinfo {author} {\bibfnamefont {F.}~\bibnamefont {Jendrzejewski}}, \bibinfo
  {author} {\bibfnamefont {M.~K.}\ \bibnamefont {Oberthaler}}, \ and\ \bibinfo
  {author} {\bibfnamefont {J.}~\bibnamefont {Berges}},\ }\bibfield  {title}
  {\enquote {\bibinfo {title} {Implementing quantum electrodynamics with
  ultracold atomic systems},}\ }\href {\doibase 10.1088/1367-2630/aa54e0}
  {\bibfield  {journal} {\bibinfo  {journal} {New Journal of Physics}\ }\textbf
  {\bibinfo {volume} {19}},\ \bibinfo {pages} {023030} (\bibinfo {year}
  {2017})}\BibitemShut {NoStop}%
\bibitem [{\citenamefont {Hauke}\ \emph {et~al.}(2013)\citenamefont {Hauke},
  \citenamefont {Marcos}, \citenamefont {Dalmonte},\ and\ \citenamefont
  {Zoller}}]{Hauke2013}%
  \BibitemOpen
  \bibfield  {author} {\bibinfo {author} {\bibfnamefont {P.}~\bibnamefont
  {Hauke}}, \bibinfo {author} {\bibfnamefont {D.}~\bibnamefont {Marcos}},
  \bibinfo {author} {\bibfnamefont {M.}~\bibnamefont {Dalmonte}}, \ and\
  \bibinfo {author} {\bibfnamefont {P.}~\bibnamefont {Zoller}},\ }\bibfield
  {title} {\enquote {\bibinfo {title} {Quantum simulation of a lattice
  {Schwinger} model in a chain of trapped ions},}\ }\href {\doibase
  10.1103/PhysRevX.3.041018} {\bibfield  {journal} {\bibinfo  {journal} {Phys.
  Rev. X}\ }\textbf {\bibinfo {volume} {3}},\ \bibinfo {pages} {041018}
  (\bibinfo {year} {2013})}\BibitemShut {NoStop}%
\bibitem [{\citenamefont {Yang}\ \emph {et~al.}(2016)\citenamefont {Yang},
  \citenamefont {Giri}, \citenamefont {Johanning}, \citenamefont {Wunderlich},
  \citenamefont {Zoller},\ and\ \citenamefont {Hauke}}]{Yang2016}%
  \BibitemOpen
  \bibfield  {author} {\bibinfo {author} {\bibfnamefont {Dayou}\ \bibnamefont
  {Yang}}, \bibinfo {author} {\bibfnamefont {Gouri~Shankar}\ \bibnamefont
  {Giri}}, \bibinfo {author} {\bibfnamefont {Michael}\ \bibnamefont
  {Johanning}}, \bibinfo {author} {\bibfnamefont {Christof}\ \bibnamefont
  {Wunderlich}}, \bibinfo {author} {\bibfnamefont {Peter}\ \bibnamefont
  {Zoller}}, \ and\ \bibinfo {author} {\bibfnamefont {Philipp}\ \bibnamefont
  {Hauke}},\ }\bibfield  {title} {\enquote {\bibinfo {title} {Analog quantum
  simulation of $(1+1)$-dimensional lattice {QED} with trapped ions},}\ }\href
  {\doibase 10.1103/PhysRevA.94.052321} {\bibfield  {journal} {\bibinfo
  {journal} {Phys. Rev. A}\ }\textbf {\bibinfo {volume} {94}},\ \bibinfo
  {pages} {052321} (\bibinfo {year} {2016})}\BibitemShut {NoStop}%
\bibitem [{\citenamefont {Mil}\ \emph {et~al.}(2020)\citenamefont {Mil},
  \citenamefont {Zache}, \citenamefont {Hegde}, \citenamefont {Xia},
  \citenamefont {Bhatt}, \citenamefont {Oberthaler}, \citenamefont {Hauke},
  \citenamefont {Berges},\ and\ \citenamefont {Jendrzejewski}}]{Mil2020}%
  \BibitemOpen
  \bibfield  {author} {\bibinfo {author} {\bibfnamefont {Alexander}\
  \bibnamefont {Mil}}, \bibinfo {author} {\bibfnamefont {Torsten~V.}\
  \bibnamefont {Zache}}, \bibinfo {author} {\bibfnamefont {Apoorva}\
  \bibnamefont {Hegde}}, \bibinfo {author} {\bibfnamefont {Andy}\ \bibnamefont
  {Xia}}, \bibinfo {author} {\bibfnamefont {Rohit~P.}\ \bibnamefont {Bhatt}},
  \bibinfo {author} {\bibfnamefont {Markus~K.}\ \bibnamefont {Oberthaler}},
  \bibinfo {author} {\bibfnamefont {Philipp}\ \bibnamefont {Hauke}}, \bibinfo
  {author} {\bibfnamefont {J{\"u}rgen}\ \bibnamefont {Berges}}, \ and\ \bibinfo
  {author} {\bibfnamefont {Fred}\ \bibnamefont {Jendrzejewski}},\ }\bibfield
  {title} {\enquote {\bibinfo {title} {A scalable realization of local {U(1)}
  gauge invariance in cold atomic mixtures},}\ }\href {\doibase
  10.1126/science.aaz5312} {\bibfield  {journal} {\bibinfo  {journal}
  {Science}\ }\textbf {\bibinfo {volume} {367}},\ \bibinfo {pages} {1128--1130}
  (\bibinfo {year} {2020})}\BibitemShut {NoStop}%
\bibitem [{\citenamefont {Yang}\ \emph {et~al.}(2020)\citenamefont {Yang},
  \citenamefont {Sun}, \citenamefont {Ott}, \citenamefont {Wang}, \citenamefont
  {Zache}, \citenamefont {Halimeh}, \citenamefont {Yuan}, \citenamefont
  {Hauke},\ and\ \citenamefont {Pan}}]{Yang2020}%
  \BibitemOpen
  \bibfield  {author} {\bibinfo {author} {\bibfnamefont {Bing}\ \bibnamefont
  {Yang}}, \bibinfo {author} {\bibfnamefont {Hui}\ \bibnamefont {Sun}},
  \bibinfo {author} {\bibfnamefont {Robert}\ \bibnamefont {Ott}}, \bibinfo
  {author} {\bibfnamefont {Han-Yi}\ \bibnamefont {Wang}}, \bibinfo {author}
  {\bibfnamefont {Torsten~V.}\ \bibnamefont {Zache}}, \bibinfo {author}
  {\bibfnamefont {Jad~C.}\ \bibnamefont {Halimeh}}, \bibinfo {author}
  {\bibfnamefont {Zhen-Sheng}\ \bibnamefont {Yuan}}, \bibinfo {author}
  {\bibfnamefont {Philipp}\ \bibnamefont {Hauke}}, \ and\ \bibinfo {author}
  {\bibfnamefont {Jian-Wei}\ \bibnamefont {Pan}},\ }\bibfield  {title}
  {\enquote {\bibinfo {title} {Observation of gauge invariance in a 71-site
  {Bose}--{Hubbard} quantum simulator},}\ }\href {\doibase
  10.1038/s41586-020-2910-8} {\bibfield  {journal} {\bibinfo  {journal}
  {Nature}\ }\textbf {\bibinfo {volume} {587}},\ \bibinfo {pages} {392--396}
  (\bibinfo {year} {2020})}\BibitemShut {NoStop}%
\bibitem [{\citenamefont {Zhou}\ \emph {et~al.}(2022)\citenamefont {Zhou},
  \citenamefont {Su}, \citenamefont {Halimeh}, \citenamefont {Ott},
  \citenamefont {Sun}, \citenamefont {Hauke}, \citenamefont {Yang},
  \citenamefont {Yuan}, \citenamefont {Berges},\ and\ \citenamefont
  {Pan}}]{Zhou2021}%
  \BibitemOpen
  \bibfield  {author} {\bibinfo {author} {\bibfnamefont {Zhao-Yu}\ \bibnamefont
  {Zhou}}, \bibinfo {author} {\bibfnamefont {Guo-Xian}\ \bibnamefont {Su}},
  \bibinfo {author} {\bibfnamefont {Jad~C.}\ \bibnamefont {Halimeh}}, \bibinfo
  {author} {\bibfnamefont {Robert}\ \bibnamefont {Ott}}, \bibinfo {author}
  {\bibfnamefont {Hui}\ \bibnamefont {Sun}}, \bibinfo {author} {\bibfnamefont
  {Philipp}\ \bibnamefont {Hauke}}, \bibinfo {author} {\bibfnamefont {Bing}\
  \bibnamefont {Yang}}, \bibinfo {author} {\bibfnamefont {Zhen-Sheng}\
  \bibnamefont {Yuan}}, \bibinfo {author} {\bibfnamefont {Jürgen}\
  \bibnamefont {Berges}}, \ and\ \bibinfo {author} {\bibfnamefont {Jian-Wei}\
  \bibnamefont {Pan}},\ }\bibfield  {title} {\enquote {\bibinfo {title}
  {Thermalization dynamics of a gauge theory on a quantum simulator},}\ }\href
  {\doibase 10.1126/science.abl6277} {\bibfield  {journal} {\bibinfo  {journal}
  {Science}\ }\textbf {\bibinfo {volume} {377}},\ \bibinfo {pages} {311--314}
  (\bibinfo {year} {2022})}\BibitemShut {NoStop}%
\bibitem [{\citenamefont {Haegeman}\ \emph {et~al.}(2011)\citenamefont
  {Haegeman}, \citenamefont {Cirac}, \citenamefont {Osborne}, \citenamefont
  {Pi\ifmmode~\check{z}\else \v{z}\fi{}orn}, \citenamefont {Verschelde},\ and\
  \citenamefont {Verstraete}}]{Haegeman2011}%
  \BibitemOpen
  \bibfield  {author} {\bibinfo {author} {\bibfnamefont {Jutho}\ \bibnamefont
  {Haegeman}}, \bibinfo {author} {\bibfnamefont {J.~Ignacio}\ \bibnamefont
  {Cirac}}, \bibinfo {author} {\bibfnamefont {Tobias~J.}\ \bibnamefont
  {Osborne}}, \bibinfo {author} {\bibfnamefont {Iztok}\ \bibnamefont
  {Pi\ifmmode~\check{z}\else \v{z}\fi{}orn}}, \bibinfo {author} {\bibfnamefont
  {Henri}\ \bibnamefont {Verschelde}}, \ and\ \bibinfo {author} {\bibfnamefont
  {Frank}\ \bibnamefont {Verstraete}},\ }\bibfield  {title} {\enquote {\bibinfo
  {title} {Time-dependent variational principle for quantum lattices},}\ }\href
  {\doibase 10.1103/PhysRevLett.107.070601} {\bibfield  {journal} {\bibinfo
  {journal} {Phys. Rev. Lett.}\ }\textbf {\bibinfo {volume} {107}},\ \bibinfo
  {pages} {070601} (\bibinfo {year} {2011})}\BibitemShut {NoStop}%
\bibitem [{\citenamefont {Haegeman}\ \emph {et~al.}(2013)\citenamefont
  {Haegeman}, \citenamefont {Osborne},\ and\ \citenamefont
  {Verstraete}}]{Haegeman2013}%
  \BibitemOpen
  \bibfield  {author} {\bibinfo {author} {\bibfnamefont {Jutho}\ \bibnamefont
  {Haegeman}}, \bibinfo {author} {\bibfnamefont {Tobias~J.}\ \bibnamefont
  {Osborne}}, \ and\ \bibinfo {author} {\bibfnamefont {Frank}\ \bibnamefont
  {Verstraete}},\ }\bibfield  {title} {\enquote {\bibinfo {title} {Post-matrix
  product state methods: To tangent space and beyond},}\ }\href {\doibase
  10.1103/PhysRevB.88.075133} {\bibfield  {journal} {\bibinfo  {journal} {Phys.
  Rev. B}\ }\textbf {\bibinfo {volume} {88}},\ \bibinfo {pages} {075133}
  (\bibinfo {year} {2013})}\BibitemShut {NoStop}%
\bibitem [{\citenamefont {Haegeman}\ \emph {et~al.}(2016)\citenamefont
  {Haegeman}, \citenamefont {Lubich}, \citenamefont {Oseledets}, \citenamefont
  {Vandereycken},\ and\ \citenamefont {Verstraete}}]{Haegeman2016}%
  \BibitemOpen
  \bibfield  {author} {\bibinfo {author} {\bibfnamefont {Jutho}\ \bibnamefont
  {Haegeman}}, \bibinfo {author} {\bibfnamefont {Christian}\ \bibnamefont
  {Lubich}}, \bibinfo {author} {\bibfnamefont {Ivan}\ \bibnamefont
  {Oseledets}}, \bibinfo {author} {\bibfnamefont {Bart}\ \bibnamefont
  {Vandereycken}}, \ and\ \bibinfo {author} {\bibfnamefont {Frank}\
  \bibnamefont {Verstraete}},\ }\bibfield  {title} {\enquote {\bibinfo {title}
  {Unifying time evolution and optimization with matrix product states},}\
  }\href {\doibase 10.1103/PhysRevB.94.165116} {\bibfield  {journal} {\bibinfo
  {journal} {Phys. Rev. B}\ }\textbf {\bibinfo {volume} {94}},\ \bibinfo
  {pages} {165116} (\bibinfo {year} {2016})}\BibitemShut {NoStop}%
\bibitem [{\citenamefont {Vanderstraeten}\ \emph {et~al.}(2019)\citenamefont
  {Vanderstraeten}, \citenamefont {Haegeman},\ and\ \citenamefont
  {Verstraete}}]{Vanderstraeten2019}%
  \BibitemOpen
  \bibfield  {author} {\bibinfo {author} {\bibfnamefont {Laurens}\ \bibnamefont
  {Vanderstraeten}}, \bibinfo {author} {\bibfnamefont {Jutho}\ \bibnamefont
  {Haegeman}}, \ and\ \bibinfo {author} {\bibfnamefont {Frank}\ \bibnamefont
  {Verstraete}},\ }\bibfield  {title} {\enquote {\bibinfo {title}
  {{Tangent-space methods for uniform matrix product states}},}\ }\href
  {\doibase 10.21468/SciPostPhysLectNotes.7} {\bibfield  {journal} {\bibinfo
  {journal} {SciPost Phys. Lect. Notes}\ ,\ \bibinfo {pages} {7}} (\bibinfo
  {year} {2019})}\BibitemShut {NoStop}%
\bibitem [{\citenamefont {Zauner}\ \emph {et~al.}(2015)\citenamefont {Zauner},
  \citenamefont {Draxler}, \citenamefont {Vanderstraeten}, \citenamefont
  {Degroote}, \citenamefont {Haegeman}, \citenamefont {Rams}, \citenamefont
  {Stojevic}, \citenamefont {Schuch},\ and\ \citenamefont
  {Verstraete}}]{Zauner2015}%
  \BibitemOpen
  \bibfield  {author} {\bibinfo {author} {\bibfnamefont {V.}~\bibnamefont
  {Zauner}}, \bibinfo {author} {\bibfnamefont {D.}~\bibnamefont {Draxler}},
  \bibinfo {author} {\bibfnamefont {L.}~\bibnamefont {Vanderstraeten}},
  \bibinfo {author} {\bibfnamefont {M.}~\bibnamefont {Degroote}}, \bibinfo
  {author} {\bibfnamefont {J.}~\bibnamefont {Haegeman}}, \bibinfo {author}
  {\bibfnamefont {M.~M.}\ \bibnamefont {Rams}}, \bibinfo {author}
  {\bibfnamefont {V.}~\bibnamefont {Stojevic}}, \bibinfo {author}
  {\bibfnamefont {N.}~\bibnamefont {Schuch}}, \ and\ \bibinfo {author}
  {\bibfnamefont {F.}~\bibnamefont {Verstraete}},\ }\bibfield  {title}
  {\enquote {\bibinfo {title} {Transfer matrices and excitations with matrix
  product states},}\ }\href {\doibase 10.1088/1367-2630/17/5/053002} {\bibfield
   {journal} {\bibinfo  {journal} {New Journal of Physics}\ }\textbf {\bibinfo
  {volume} {17}},\ \bibinfo {pages} {053002} (\bibinfo {year}
  {2015})}\BibitemShut {NoStop}%
\bibitem [{\citenamefont {Zauner-Stauber}\ and\ \citenamefont
  {Halimeh}(2017)}]{Zauner2017}%
  \BibitemOpen
  \bibfield  {author} {\bibinfo {author} {\bibfnamefont {Valentin}\
  \bibnamefont {Zauner-Stauber}}\ and\ \bibinfo {author} {\bibfnamefont
  {Jad~C.}\ \bibnamefont {Halimeh}},\ }\bibfield  {title} {\enquote {\bibinfo
  {title} {Probing the anomalous dynamical phase in long-range quantum spin
  chains through {Fisher}-zero lines},}\ }\href {\doibase
  10.1103/PhysRevE.96.062118} {\bibfield  {journal} {\bibinfo  {journal} {Phys.
  Rev. E}\ }\textbf {\bibinfo {volume} {96}},\ \bibinfo {pages} {062118}
  (\bibinfo {year} {2017})}\BibitemShut {NoStop}%
\bibitem [{\citenamefont {Pedersen}\ and\ \citenamefont
  {Zinner}(2021)}]{Pedersen2021}%
  \BibitemOpen
  \bibfield  {author} {\bibinfo {author} {\bibfnamefont {Simon~Panyella}\
  \bibnamefont {Pedersen}}\ and\ \bibinfo {author} {\bibfnamefont
  {Nikolaj~Thomas}\ \bibnamefont {Zinner}},\ }\bibfield  {title} {\enquote
  {\bibinfo {title} {Lattice gauge theory and dynamical quantum phase
  transitions using noisy intermediate-scale quantum devices},}\ }\href
  {\doibase 10.1103/PhysRevB.103.235103} {\bibfield  {journal} {\bibinfo
  {journal} {Phys. Rev. B}\ }\textbf {\bibinfo {volume} {103}},\ \bibinfo
  {pages} {235103} (\bibinfo {year} {2021})}\BibitemShut {NoStop}%
\bibitem [{\citenamefont {Jensen}\ \emph {et~al.}(2022)\citenamefont {Jensen},
  \citenamefont {Pedersen},\ and\ \citenamefont {Zinner}}]{Jensen2022}%
  \BibitemOpen
  \bibfield  {author} {\bibinfo {author} {\bibfnamefont {Rasmus~Berg}\
  \bibnamefont {Jensen}}, \bibinfo {author} {\bibfnamefont {Simon~Panyella}\
  \bibnamefont {Pedersen}}, \ and\ \bibinfo {author} {\bibfnamefont
  {Nikolaj~Thomas}\ \bibnamefont {Zinner}},\ }\bibfield  {title} {\enquote
  {\bibinfo {title} {Dynamical quantum phase transitions in a noisy lattice
  gauge theory},}\ }\href {\doibase 10.1103/PhysRevB.105.224309} {\bibfield
  {journal} {\bibinfo  {journal} {Phys. Rev. B}\ }\textbf {\bibinfo {volume}
  {105}},\ \bibinfo {pages} {224309} (\bibinfo {year} {2022})}\BibitemShut
  {NoStop}%
\bibitem [{\citenamefont {{Van Damme}}\ \emph {et~al.}(2022)\citenamefont {{Van
  Damme}}, \citenamefont {{Zache}}, \citenamefont {{Banerjee}}, \citenamefont
  {{Hauke}},\ and\ \citenamefont {{Halimeh}}}]{VanDamme2022}%
  \BibitemOpen
  \bibfield  {author} {\bibinfo {author} {\bibfnamefont {Maarten}\ \bibnamefont
  {{Van Damme}}}, \bibinfo {author} {\bibfnamefont {Torsten~V.}\ \bibnamefont
  {{Zache}}}, \bibinfo {author} {\bibfnamefont {Debasish}\ \bibnamefont
  {{Banerjee}}}, \bibinfo {author} {\bibfnamefont {Philipp}\ \bibnamefont
  {{Hauke}}}, \ and\ \bibinfo {author} {\bibfnamefont {Jad~C.}\ \bibnamefont
  {{Halimeh}}},\ }\bibfield  {title} {\enquote {\bibinfo {title} {{Dynamical
  quantum phase transitions in spin-$S$ $\mathrm{U}(1)$ quantum link
  models}},}\ }\href@noop {} {\bibfield  {journal} {\bibinfo  {journal} {arXiv
  e-prints}\ } (\bibinfo {year} {2022})},\ \Eprint
  {http://arxiv.org/abs/2203.01337} {arXiv:2203.01337 [cond-mat.str-el]}
  \BibitemShut {NoStop}%
\bibitem [{\citenamefont {Mueller}\ \emph {et~al.}(2022)\citenamefont
  {Mueller}, \citenamefont {Carolan}, \citenamefont {Connelly}, \citenamefont
  {Davoudi}, \citenamefont {Dumitrescu},\ and\ \citenamefont
  {Yeter-Aydeniz}}]{Mueller2022DQPT}%
  \BibitemOpen
  \bibfield  {author} {\bibinfo {author} {\bibfnamefont {Niklas}\ \bibnamefont
  {Mueller}}, \bibinfo {author} {\bibfnamefont {Joseph~A.}\ \bibnamefont
  {Carolan}}, \bibinfo {author} {\bibfnamefont {Andrew}\ \bibnamefont
  {Connelly}}, \bibinfo {author} {\bibfnamefont {Zohreh}\ \bibnamefont
  {Davoudi}}, \bibinfo {author} {\bibfnamefont {Eugene~F.}\ \bibnamefont
  {Dumitrescu}}, \ and\ \bibinfo {author} {\bibfnamefont {Kübra}\ \bibnamefont
  {Yeter-Aydeniz}},\ }\bibfield  {title} {\enquote {\bibinfo {title} {Quantum
  computation of dynamical quantum phase transitions and entanglement
  tomography in a lattice gauge theory},}\ }\href {\doibase
  10.48550/ARXIV.2210.03089} {\  (\bibinfo {year} {2022}),\
  10.48550/ARXIV.2210.03089}\BibitemShut {NoStop}%
\bibitem [{\citenamefont {Jurcevic}\ \emph {et~al.}(2017)\citenamefont
  {Jurcevic}, \citenamefont {Shen}, \citenamefont {Hauke}, \citenamefont
  {Maier}, \citenamefont {Brydges}, \citenamefont {Hempel}, \citenamefont
  {Lanyon}, \citenamefont {Heyl}, \citenamefont {Blatt},\ and\ \citenamefont
  {Roos}}]{Jurcevic2017}%
  \BibitemOpen
  \bibfield  {author} {\bibinfo {author} {\bibfnamefont {P.}~\bibnamefont
  {Jurcevic}}, \bibinfo {author} {\bibfnamefont {H.}~\bibnamefont {Shen}},
  \bibinfo {author} {\bibfnamefont {P.}~\bibnamefont {Hauke}}, \bibinfo
  {author} {\bibfnamefont {C.}~\bibnamefont {Maier}}, \bibinfo {author}
  {\bibfnamefont {T.}~\bibnamefont {Brydges}}, \bibinfo {author} {\bibfnamefont
  {C.}~\bibnamefont {Hempel}}, \bibinfo {author} {\bibfnamefont {B.~P.}\
  \bibnamefont {Lanyon}}, \bibinfo {author} {\bibfnamefont {M.}~\bibnamefont
  {Heyl}}, \bibinfo {author} {\bibfnamefont {R.}~\bibnamefont {Blatt}}, \ and\
  \bibinfo {author} {\bibfnamefont {C.~F.}\ \bibnamefont {Roos}},\ }\bibfield
  {title} {\enquote {\bibinfo {title} {Direct observation of dynamical quantum
  phase transitions in an interacting many-body system},}\ }\href {\doibase
  10.1103/PhysRevLett.119.080501} {\bibfield  {journal} {\bibinfo  {journal}
  {Phys. Rev. Lett.}\ }\textbf {\bibinfo {volume} {119}},\ \bibinfo {pages}
  {080501} (\bibinfo {year} {2017})}\BibitemShut {NoStop}%
\bibitem [{\citenamefont {Coleman}(1976)}]{Coleman1976}%
  \BibitemOpen
  \bibfield  {author} {\bibinfo {author} {\bibfnamefont {Sidney}\ \bibnamefont
  {Coleman}},\ }\bibfield  {title} {\enquote {\bibinfo {title} {More about the
  massive {Schwinger} model},}\ }\href {\doibase
  https://doi.org/10.1016/0003-4916(76)90280-3} {\bibfield  {journal} {\bibinfo
   {journal} {Annals of Physics}\ }\textbf {\bibinfo {volume} {101}},\ \bibinfo
  {pages} {239 -- 267} (\bibinfo {year} {1976})}\BibitemShut {NoStop}%
\bibitem [{\citenamefont {{Desaules}}\ \emph
  {et~al.}(2022{\natexlab{b}})\citenamefont {{Desaules}}, \citenamefont
  {{Hudomal}}, \citenamefont {{Banerjee}}, \citenamefont {{Sen}}, \citenamefont
  {{Papi{\'c}}},\ and\ \citenamefont {{Halimeh}}}]{Desaules2022b}%
  \BibitemOpen
  \bibfield  {author} {\bibinfo {author} {\bibfnamefont {Jean-Yves}\
  \bibnamefont {{Desaules}}}, \bibinfo {author} {\bibfnamefont {Ana}\
  \bibnamefont {{Hudomal}}}, \bibinfo {author} {\bibfnamefont {Debasish}\
  \bibnamefont {{Banerjee}}}, \bibinfo {author} {\bibfnamefont {Arnab}\
  \bibnamefont {{Sen}}}, \bibinfo {author} {\bibfnamefont {Zlatko}\
  \bibnamefont {{Papi{\'c}}}}, \ and\ \bibinfo {author} {\bibfnamefont
  {Jad~C.}\ \bibnamefont {{Halimeh}}},\ }\bibfield  {title} {\enquote {\bibinfo
  {title} {Prominent quantum many-body scars in a truncated {Schwinger}
  model},}\ }\href@noop {} {\bibfield  {journal} {\bibinfo  {journal} {arXiv
  e-prints}\ ,\ \bibinfo {eid} {arXiv:2204.01745}} (\bibinfo {year}
  {2022}{\natexlab{b}})},\ \Eprint {http://arxiv.org/abs/2204.01745}
  {arXiv:2204.01745 [cond-mat.quant-gas]} \BibitemShut {NoStop}%
\end{thebibliography}%
\end{document}